\documentclass[10pt,conference]{IEEEtran}
\usepackage{cite}
\usepackage{amsmath,amssymb,amsfonts}
\usepackage{algorithmic}
\usepackage{graphicx}
\usepackage{textcomp}
\usepackage{xcolor}
\usepackage[hyphens]{url}
\UseRawInputEncoding

\def\BibTeX{{\rm B\kern-.05em{\sc i\kern-.025em b}\kern-.08em
    T\kern-.1667em\lower.7ex\hbox{E}\kern-.125emX}}

\newcommand{\amshbar}{\hbar}
\usepackage{tikz}
\usepackage{quantikz}
\usepackage{listings}
\usepackage{caption}
\usepackage{float}
\usepackage{placeins}

\lstnewenvironment{pythoncode}{\lstset{language=Python}}{}

\title{Fidelity estimator, randomized benchmarking and ZNE for quantum pulses} 
\author{
    \IEEEauthorblockN{
    Jinglei Cheng\IEEEauthorrefmark{1} \ \
    Zhiding Liang\IEEEauthorrefmark{2} \ \
    Rui Yang\IEEEauthorrefmark{3} \ \
    Hang Ren\IEEEauthorrefmark{4} \ \
    Yiyu Shi \IEEEauthorrefmark{2}
    Tongyang Li\IEEEauthorrefmark{3} \ \   
    Xuehai Qian\IEEEauthorrefmark{1}
    } \ \
    \IEEEauthorblockA{
    \IEEEauthorrefmark{1}Purdue University \ 
    \IEEEauthorrefmark{2}University of Notre Dame \
    }
    \IEEEauthorblockA{
    \IEEEauthorrefmark{3}Peking University \ 
    \IEEEauthorrefmark{4}University of California, Berkeley \ 
    }

}

\begin{document}
\maketitle
\thispagestyle{plain}
\pagestyle{plain}

\begin{abstract}
    
%%%%%% -- PAPER CONTENT STARTS-- %%%%%%%%

Quantum pulses and quantum gates are both important components in quantum computing, but they operate at different abstraction layers. 
Quantum pulses often refer to the physical control signals applied to manipulate qubits, while quantum gates are abstract mathematical representations which are usually unitary operations on qubits.
When quantum programs are synthesized into quantum gates and executed on quantum devices, these gates are eventually transformed into physical control signals. 
In this sense, quantum gates can be considered as groups of well-calibrated quantum pulses. 
Adopting quantum pulses directly in quantum programs, particularly in variational quantum algorithms, can bring benefits such as reduced circuit latency.
% \jlc{Be more specific about the reason of pulse advantage in VQA.}
This is because variational quantum algorithms do not demand high-precision operations.
Another reason for the increasing attention in quantum pulses is the capability of implementing operations that are not achievable by quantum gates. 
These operations include accessing higher energy states, directly implementing unitary matrices, and introducing entanglement of qubits beyond CX or iSWAP gates.

% At present, few studies have explored the profiling of quantum pulses. 
Most previous research focused on designing pulse programs without considering the performance of individual elements or the final fidelity. 
To evaluate the performance of quantum pulses, it is required to know the noiseless results of the pulses.
However, quantum pulses can implement unitary matrices that are not analytically known to the user, and pulse simulator usually comes with significant computational overhead.
Consequently, determining fidelity of a pulse program is challenging without the knowledge of the ideal results.
In this paper, we propose to use reversed pulses to evaluate the performance of quantum pulses, which can provide guidance to design pulse programs. 
By employing reversed pulses, we can ensure that, in the noiseless situation, the final quantum states are the same as the initial states. 
This method enables us to evaluate the fidelity of pulse programs by measuring the difference between the final states and the initial states.
% Lack of fidelity estimator? (why need?)
% The method can also be adopted as fidelity estimators for pulse programs. 
Such fidelity estimator can tell whether the results are meaningful for quantum pulses on real quantum machines. 
% With our proposed reverse pulses, we can create reversed pulse programs, enabling us to evaluate the fidelity of the original pulse program.
% Lack of QEC on pulses ? (why need ?)
There are various quantum error correction (QEC) methods available for gate circuits; however, few studies have demonstrated QEC on pulse-level programs. 
% QEC techniques can improve the performance of quantum pulses. \ry{error mitigation?}
In this paper, we use reversed pulses to implement zero noise extrapolation (ZNE) on pulse programs and demonstrate results for variational quantum eigensolver (VQE) tasks. 
% \jlc{Fill in the experimental resutls pls.}
% We achieve an accuracy of 0 percent for the hydrogen molecule ($H_2$) as a demonstration of this approach. 
The deviation from the idea energy value is reduced by an average of 54.1\% with our techniques.
% \tyl{abstract too long?}

% Our proposed solution ? (key insight: use reverse pulse)

% Our results ? (pulse set profiling results, fidelity estimator, ZNE)

\end{abstract}

% % \newpage

\section{Introduction}

% \jlc{What is current progress of quantum computing?}

Quantum computing is a rapidly developing field at the intersection of physics, computer science, and mathematics. This interdisciplinary area of research explores the utilization of quantum mechanical phenomena, such as superposition and entanglement, to perform calculations and solve problems that are either intractable or computationally expensive for classical computers~\cite{preskill2018quantum}. The fundamental building block of quantum computing is the quantum bit or qubit, which, in contrast to classical bits that can only represent a binary state (0 or 1), can exist in a superposition of states, which enables the simultaneous processing of large amounts of information~\cite{liang2021can}.

Current research in quantum computing witnesses both hardware- and software-level advancements. On the hardware side, several physical implementations of qubits, including superconducting circuits~\cite{krantz2019quantum, kjaergaard2020superconducting}, trapped ions~\cite{cirac1995quantum, leibfried2003quantum, blatt2012quantum}, quantum dots~\cite{imamog1999quantum, petta2005coherent, englund2005controlling, hanson2007spins}, ultracold atoms~\cite{jaksch2005cold, lewenstein2007ultracold, bloch2008many, gross2017quantum}, are being explored to develop scalable and fault-tolerant quantum computers. 
% The development of error-correcting codes and noise-reduction techniques aims to mitigate the noise of quantum states due to various errors including decoherence. 
Software-level research is dedicated to designing quantum algorithms and optimization for compiling and simulating quantum circuits on classical hardware.

\begin{figure}
    \centering
    \includegraphics[width=0.5\textwidth]{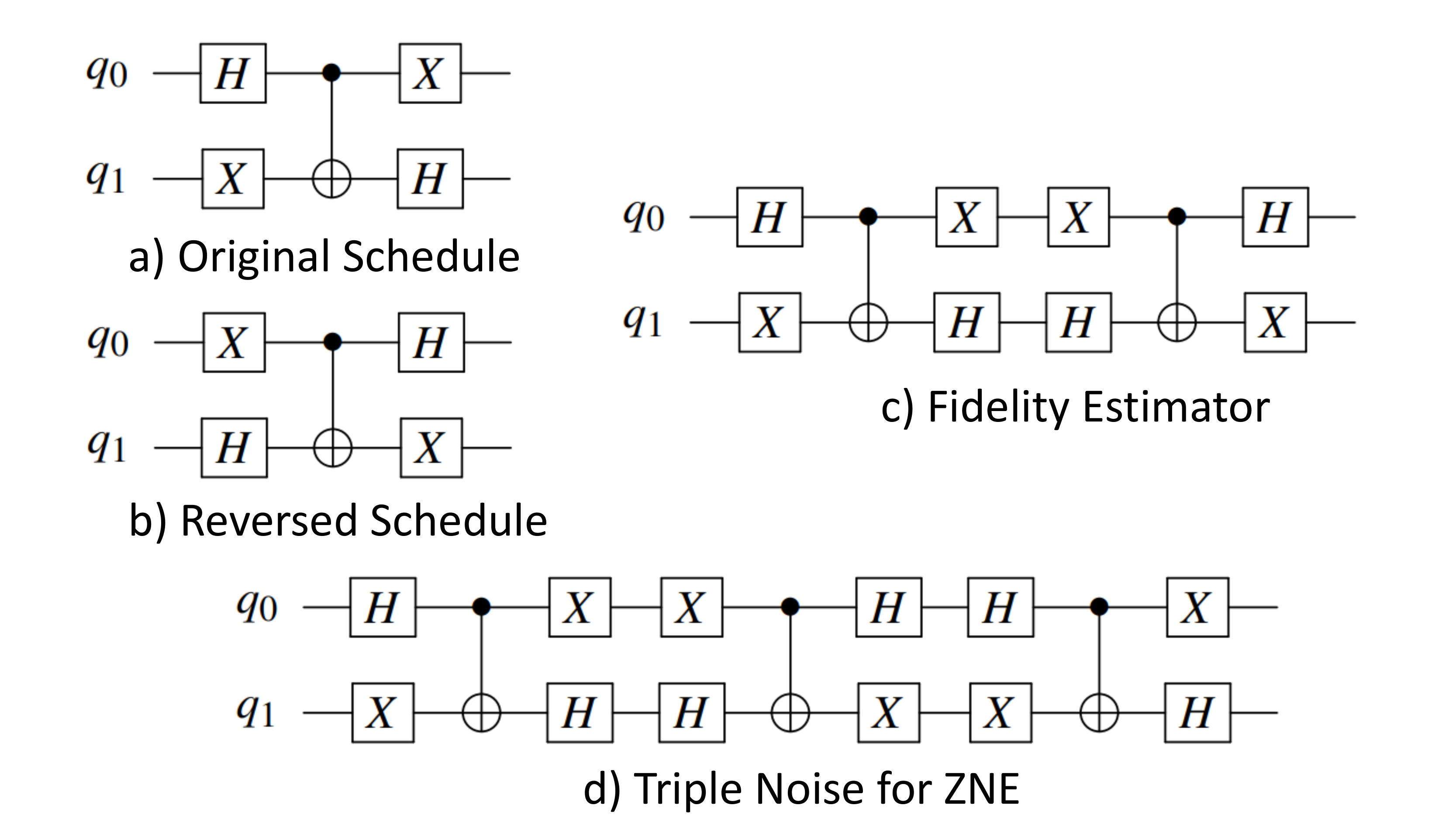}
    \caption{Example of the reversed circuit, fidelity estimator and triple noise circuit for ZNE on the gate-level. To measure the fidelity of the original schedule, we need to conduct the state tomography and compare it with the measurement outcomes. The process is computationally expensive. Therefore, we can append the reversed circuit to the original circuit and approximate the true fidelity by the probability of successful trials (PST). If we append another original circuit to the fidelity estimator, we obtain a circuit that has triple noise and the same function. The triple-noise circuit is one way to expand noise in the ZNE.}
    \label{fig:ZNE_demo}
\end{figure}

% \jlc{Why people use quantum pulses?(potentials of quantum pulses)}

% Quantum gates can be viewed as groups of pulses that are regularly calibrated to ensure high fidelity operations. The gate-level abstraction layer provides a well-defined set of operations that can be used to construct quantum circuits for implementing quantum algorithms. However, such sets of pulses often use redundancy to ensure controllability and accuracy~\cite{}, leading to longer circuit latencies and increased decoherence errors.
% In variational quantum algorithms (VQAs), circuit latency can be reduced by waiving redundancy, allowing for more flexibility and expressibility. For example, parameterized pulses can replace parameterized gates, and the variational feature of the algorithm can automatically adjust for errors including underrotation and overrotation. Also, as recent works have pointed out, techniques such as quantum optimal control~\cite{choquette2021quantum, meitei2021gate} can be adopted to significantly shorten the circuit latency.
% Overall, pulse-level abstraction layer provides researchers with the benefits of shorter circuit latency and more flexibility.

% \jlc{Examples of quantum pulses?}

\begin{figure}[htb]
    \centering
    \includegraphics[width=0.5\textwidth]{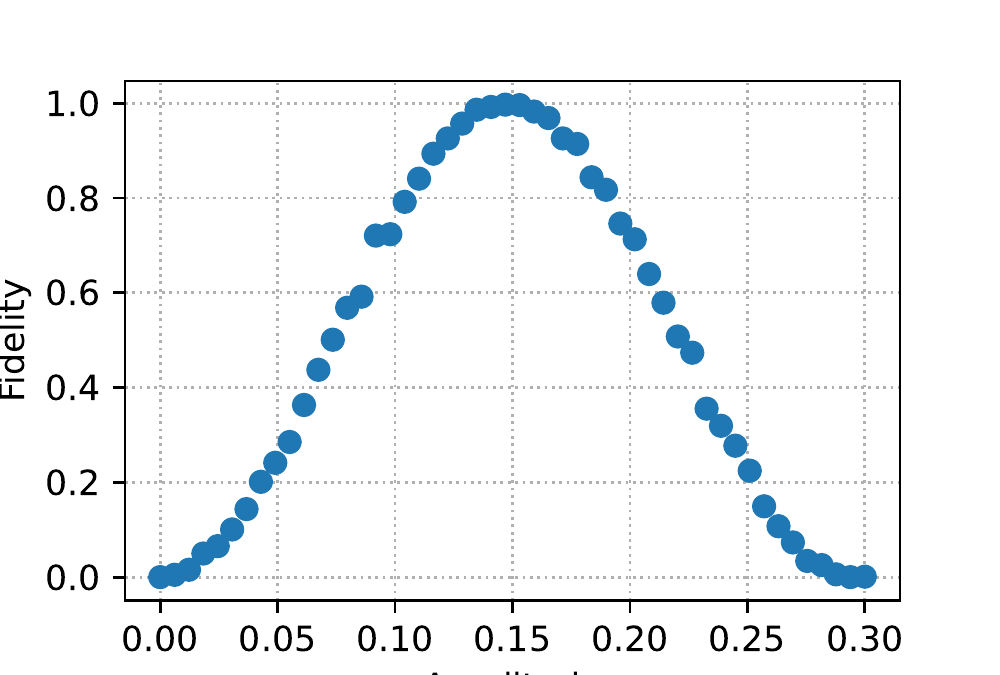}
    \caption{The amplitude calibration of X gate. In gate-level compilation, calibration is needed for single-qubit gates. If the X gate is implemented with $DRAG$ pulse, parameters of $DRAG$ pulse also need to be tuned to minimize the energy leakage and phase error~\cite{motzoi2009simple}.}
    \label{fig:XCali}
\end{figure}

Qiskit~\cite{mckay2018qiskit, aleksandrowicz2019qiskit} is one of the most popular platforms for quantum computing. It currently offers developers access to pulse-level controls through Qiskit Pulse~\cite{alexander2020qiskit}. This framework is flexible and can be extended to design and generate quantum pulses. One commonly used single-qubit pulse is the $DRAG$ Pulse~\cite{krantz2019quantum, motzoi2009simple, gambetta2011analytic, de2015fast}, which is characterized by parameters such as duration, amplitude, $\sigma$, and $\beta$. These parameters determine the shape and the function of the pulse. The $DRAG$ Pulse is often calibrated to implement rotations around the X-axis, minimizing leakage and phase errors, as shown in Figure~\ref{fig:XCali}. In addition to single-qubit pulses, Qiskit Pulse also supports multi-qubit pulses such as the Cross Resonance ($CR$) pulse~\cite{alexander2020qiskit, krantz2019quantum, de2010selective, rigetti2010fully}. This pulse generates entanglement between a control qubit and a target qubit. The $CR$ pulse assumes a Cross Resonance Hamiltonian with a specific mathematical form:
\begin{equation}
\begin{aligned}
H &= Z \otimes A_2 + I \otimes B_2 \nonumber \\
&= a_x \hat{Z} \hat{X} + a_y \hat{Z} \hat{Y} + a_z \hat{Z} \hat{Z} + b_x \hat{I} \hat{X} + b_y \hat{I} \hat{Y} + b_z \hat{I} \hat{Z}.
\end{aligned}
\end{equation}
% The entangling operation is achieved by the $ZX$ term. 
Usually the $CR$ pulse is calibrated to suppress interaction terms other than $ZX$ to implement $R_{ZX}(\pi/4)$, which can be used to implement CNOT gate.
To calibrate the $CR$ pulse and mitigate unwanted interactions, methods like echoed $CR$ and phase offset can be employed~\cite{earnest2021pulse}.
$DRAG$ and $CR$ are the simplest pulses that can implement a universal set of operations.

% Pulse Channels:
% \begin{itemize}
%     \item DriveChannel:
%     \item ControlChannel:
%     \item MeasureChannel:
%     \item AcquireChannel:
% \end{itemize}

% Typical operations: 
% \begin{itemize}
%     \item Play:
%     \item Delay:
%     \item ShiftPhase:
%     \item SetFrequency:
% \end{itemize}

% \jlc{Defects of quantum pulses?}

While quantum pulses offer several advantages over quantum gates, they also have notable drawbacks that must be addressed~\cite{smith2022summary}. One significant issue is that pulse program design is heavily dependent on the underlying hardware properties of quantum devices. This means that pulse programs must be designed for each specific quantum computer, and they will produce very different results on different quantum hardware.
Moreover, since the pulses directly interact with qubits, there are constraints on their form. For instance, pulse programs may suffer from energy leakage~\cite{motzoi2009simple}, high-energy state access, and cross-talk between different qubits~\cite{sheldon2016procedure}. These problems are often hidden behind the gate-level abstraction layer, but they can significantly impact the performance of pulse programs.
% Developers need to carefully consider the hardware properties and constraints when designing pulse programs, which can be a challenging task. Nonetheless, addressing these issues can lead to further optimization and better performance of quantum pulses.

% \jlc{Make defects of pulse simulator shorter to resolve the problem of overlapp between motivation}
% \jlc{Defects of pulse simulator?}
When experimenting with quantum pulses, pulse simulators also have several drawbacks and limitations. In a pulse simulator, system hamiltonians derived from real quantum machines are used to model the behavior of pulses on backends~\cite{Li2022pulselevelnoisy}. However, these system hamiltonians may not fully capture the intricate and time-varying nature of real quantum hardware, making it difficult to model accurately. Consequently, simulation results may not match the behavior of quantum hardware, particularly when it comes to characterizing noise and errors.
Furthermore, pulse simulators are typically time-expensive and precision-limited. This is because they simulate the behavior of quantum pulses as if they are simulating quantum gates, which means that each time slot of pulses corresponds to a gate, resulting in a massive number of operations that need to be simulated even in a short pulse circuit.
Overall, current pulse simulators are expensive in terms of time and not accurate enough for evaluating certain applications.

% \jlc{Our proposed technique, contributions, results.}

Therefore, in this paper, we propose to take the advantage of reversed quantum pulses to evaluate the fidelity of pulse programs and perform quantum error mitigation. reversed pulses refer to the pulses that can undo the function of a quantum pulse. For example, if a quantum pulse implements the function of $R_x(\pi/3)$, then the reversed pulse should implement a $R_x(-\pi/3)$. We can revert the quantum state back to its original state with reversed pulse circuits. In this context, the noise of the final circuits, which implement the identity matrix, corresponds proportionally to the noise in the original pulse circuits for which we want to determine the fidelity. By comparing the final state with the initial state, we can devise a fidelity estimator for pulse circuits. 

In this study, we also assess the fidelities for pulses of different shapes, the results of which may help the design of pulse circuits. 
When we evaluate the fidelities for pulses of different shapes, we use a circuit of repeating pulses with random parameters. Then, the reversed pulse circuit is generated and appended to the end. The results reveal great differences in noise levels for pulses of different shapes. For pulses of favorable shapes, the infidelity increases gradually as the number of pulses in the sequence grows, while maintaining reasonably small error bars. In contrast, for pulses of ``bad'' shapes, the infidelity increases rapidly with the number of pulses, and error bars are much longer in the figure, as shown in Figure~\ref{fig:output}. 

The errors in quantum computing can be roughly divided into these components: control errors that are caused by inaccurate operations, coherent and decoherent errors that are determined by the circuit duration and the measurement errors that are caused by the readout of the quantum information~\cite{larose2022mitiq}.
~\cite{kandala2019error} tries to mitigate the time-dependent errors by stretching the duration of the circuits. We note that the stretched circuit can indicate the quantity of time-dependent errors, but it doesn't capture the control errors in the original circuit. For example, when the circuit is stretched to be 50\% longer than its original duration. The control errors won't proportionally grow by 50\%. If the pulse operations remain unchanged and additional time-delay slots are added into the circuit, the control errors stay the same. However, if the pulse operations are stretched to have greater latency and lower amplitudes, the control errors will not be exactly 1.5 times their original value. 
In this example, the ``nonlinearity'' of the noise with respect to the noise scale number will make it difficult to effectively encounter control errors.
Therefore, in this paper, we propose to use folded pulse circuits to increase the noise level and use zero noise extrapolation (ZNE) ~\cite{kandala2019error,giurgica2020digital} to conduct error mitigation. For a given pulse, the reversed pulse is created with identical amplitude and ``reversed'' phase. In this way, both control errors and time-dependent errors are simultaneously magnified. The types of errors that our method overlooks are measurement errors, which can be mitigated with orthogonal techniques~\cite{barron2020measurement}.

% \tyl{quantum noise mitigation?}
% \tyl{a bit more introduction here: what's reverse pulses?}
The main contributions of our work are:
\begin{enumerate}
    \item Fidelity estimator: we create a fidelity estimator for quantum pulses with reversed pulses. We demonstrate a reliable method to reverse pulse programs. 
    \item Randomized benchmarking: we perform randomized benchmarking for parameterized pulses of different shapes with our fidelity estimator. Results show significantly different errors for different pulses. $DRAG$ pulse has the lowest infidelity per pulse of approximately 0.04\%, while the $Square$ pulse the highest infidelity per pulse of approximately 1\% per pulse. All the pulses have the same duration of 160 $dt$ or 35.5 $ns$.
    \item Zero noise extrapolation: we adopt reversed pulse to implement ZNE on VQE tasks. The deviation from the idea energy value is reduced by an average of 54.1\% with our techniques.
    
\end{enumerate}
The results of our study indicate that reversed pulse is effective on both real machines and simulators, which can provide information on the properties of parameterized pulses. 
% Additionally, we use ZNE to implement a pulse-level VQE example with reduced deviation from the ideal value.

% \tyl{summarize our results}

% \begin{equation*}
%     \begin{aligned}
%         H/ \amshbar \ & =  \ \sum_{i=0}^{4} \ \left( \ \frac{ \ \omega_{q,i}}{2}( \ \mathbb{I}- \ \sigma_i^{z})+ \ \frac{ \ \Delta_{i}}{2}(O_i^2-O_i)+ \ \Omega_{d,i}D_i(t) \ \sigma_i^{X} \ \right)  \ \ \\ \ & + J_{0,1}( \ \sigma_{0}^{+} \ \sigma_{1}^{-}+ \ \sigma_{0}^{-} \ \sigma_{1}^{+}) + J_{1,2}( \ \sigma_{1}^{+} \ \sigma_{2}^{-}+ \ \sigma_{1}^{-} \ \sigma_{2}^{+}) \\ & + J_{2,3}( \ \sigma_{2}^{+} \ \sigma_{3}^{-}+ \ \sigma_{2}^{-} \ \sigma_{3}^{+}) + J_{3,4}( \ \sigma_{3}^{+} \ \sigma_{4}^{-}+ \ \sigma_{3}^{-} \ \sigma_{4}^{+})  \ \ \\ \ & +  \ \Omega_{d,0}(U_{0}^{(0,1)}(t)) \ \sigma_{0}^{X} +  \ \Omega_{d,1}(U_{1}^{(1,0)}(t)+U_{2}^{(1,2)}(t)) \ \sigma_{1}^{X}  \ \\ \ \ & +  \ \Omega_{d,2}(U_{3}^{(2,1)}(t)+U_{4}^{(2,3)}(t)) \ \sigma_{2}^{X} \ \\ & +  \ \Omega_{d,3}(U_{6}^{(3,4)}(t)+U_{5}^{(3,2)}(t)) \ \sigma_{3}^{X}  \ \  +  \ \Omega_{d,4}(U_{7}^{(4,3)}(t)) \ \sigma_{4}^{X}
%     \end{aligned}
% \end{equation*}

% \clearpage

\section{Background}

\subsection{Pulses in superconducting quantum computers}

\begin{figure}[htb]
    \centering
    \includegraphics[width=0.5\textwidth]{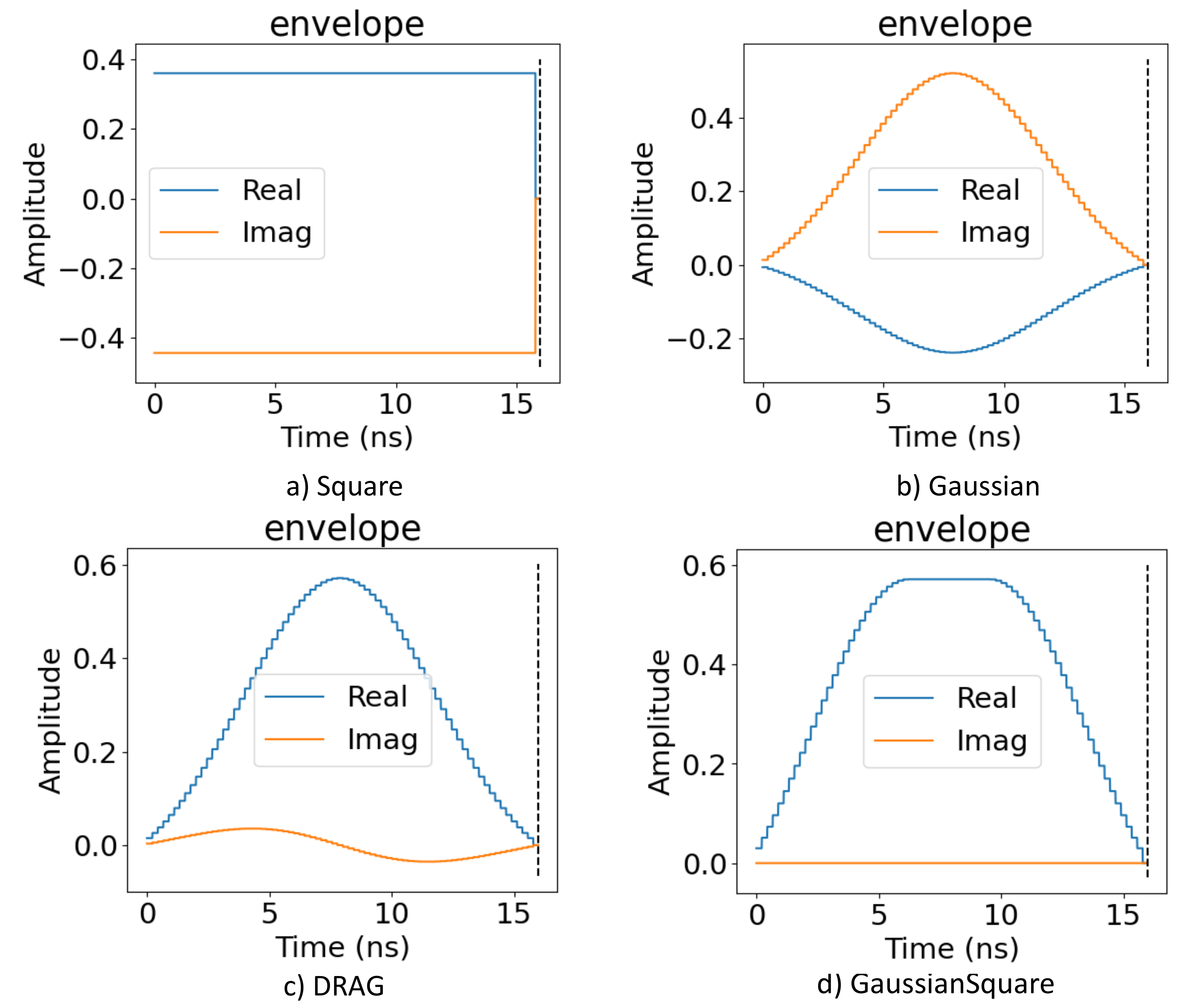}
    \caption{Different shapes of parameterized pulses. These pulses have common parameters including amplitude, angle. $DRAG$ and $Gaussian$ pulses have more parameters specifying the shapes. The real and imaginary components will be combined and modulate a high-frequency signal within a IQ-mixer. }
    \label{pulseshape}
\end{figure}

\begin{figure}[htb]
    \centering
    \includegraphics[width=0.5\textwidth]{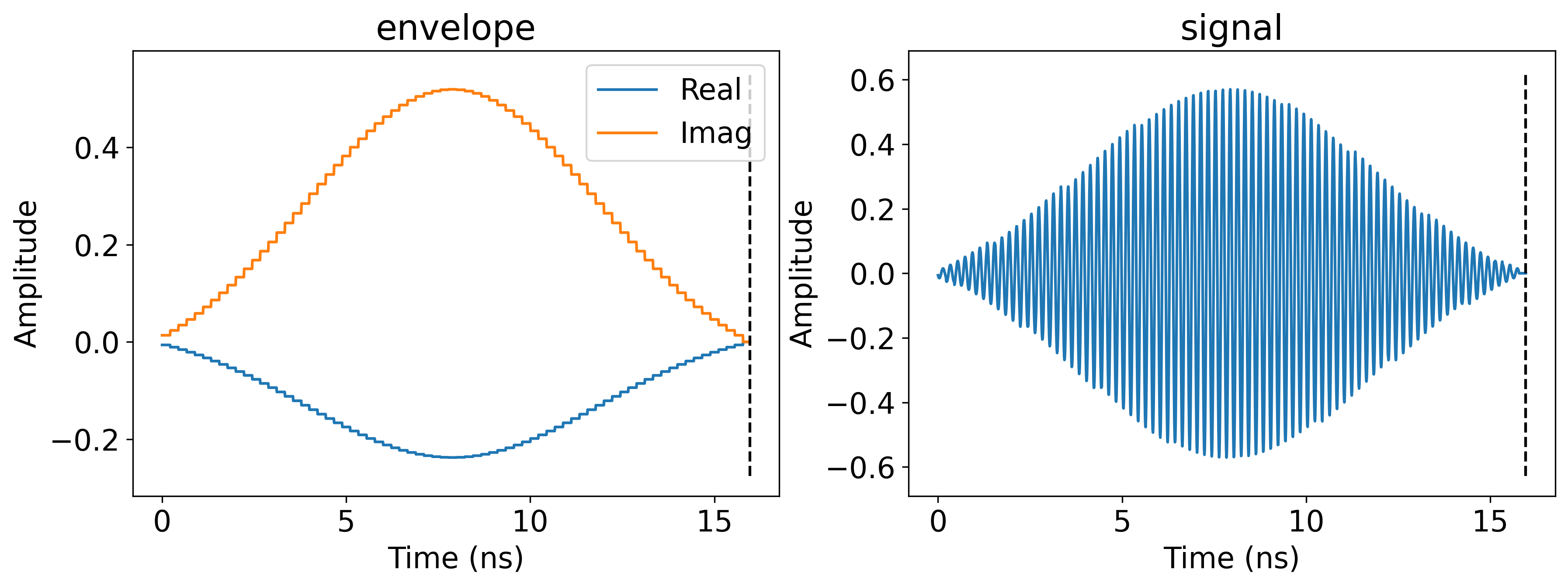}
    \caption{In quantum pulse programs, the parameterized pulses that we fine are envelopes of high-frequency microwave carrier signal. Shown in the figure is an exmaple of $Gaussian$ pulses and the corresponding signal that will be sent to transmon circuit.}
    \label{carrier signals}
\end{figure}

% Background takes 1 page

% How pulses work of superconducting quantum computer?

% Equations of mixer? AWG? System Hamiltonian? (how our pulses per dt change the state of qubit?)

% How to design pulses ? (energy leakage($DRAG$), cross-talk mitigation($CR$), CNOT design($CR$))

% Current pulse simulator ?

% Current pulse composer (qiskit)?

% Current hardware support of pulses? (qiskit)

% %Qiskit pulse related background
% Refer to <<A quantum engineer's guide to superconducting qubits>>

Superconducting quantum computers use transmons as qubits.
A microwave source provides a high-frequency signal, and an arbitrary waveform generator (AWG) delivers a pulse envelope.
% which is occasionally accompanied by a low-frequency component. 
The IQ-mixer then merges these signals to produce a shaped waveform at a frequency that usually resonates with the qubit~\cite{krantz2019quantum, ryan2017hardware}.

% The Hamiltonian square represents the entire energy of the quantum system, including the interactions between qubits. 
The time-dependent Hamiltonian equation H(t) describes the evolution of qubits. The Hamiltonian equation for a single qubit driven by a microwave pulse can be written as

% \begin{equation}
$$
    H(t)/ \amshbar = - \frac{\omega_q}{2}\sigma_z -  \frac{\Omega(t)}{2}\sigma_x
$$
% \end{equation}
where $\amshbar$ is the reduced Planck constant, $\omega_q$ is the qubit's frequency, $\sigma_x$ and $\sigma_z$ are Pauli matrices, and $\Omega(t)$ reflects the strength of control pulses. Tuning of $\Omega(t)$ allows for the manipulation of qubit states.
% When designing pulse programs for superconducting quantum computers, we need to consider factors such as energy leakage, crosstalk to maximize the pulse fidelity. 
% The design process aims to determine the shape, duration, amplitude and phase for quantum pulses. 
% \tyl{how can those factors impact the above Hamiltonian?} 

% The pulse shape, parameters, and control techniques are all critical components of the design process. 

% The pulse design discussed in this article is based on the Qiskit platform, which is called Qiskit Pulse \tyl{add citation}, a library module within the platform, that allows users to construct and manipulate pulse schedules tailored to quantum circuits. This capability enables the creation of custom pulse shapes, the application of these shapes to designated qubits, and the development of intricate pulse sequences to facilitate quantum operations.

\begin{figure}[htb]
    \centering
    \includegraphics[width=0.5\textwidth]{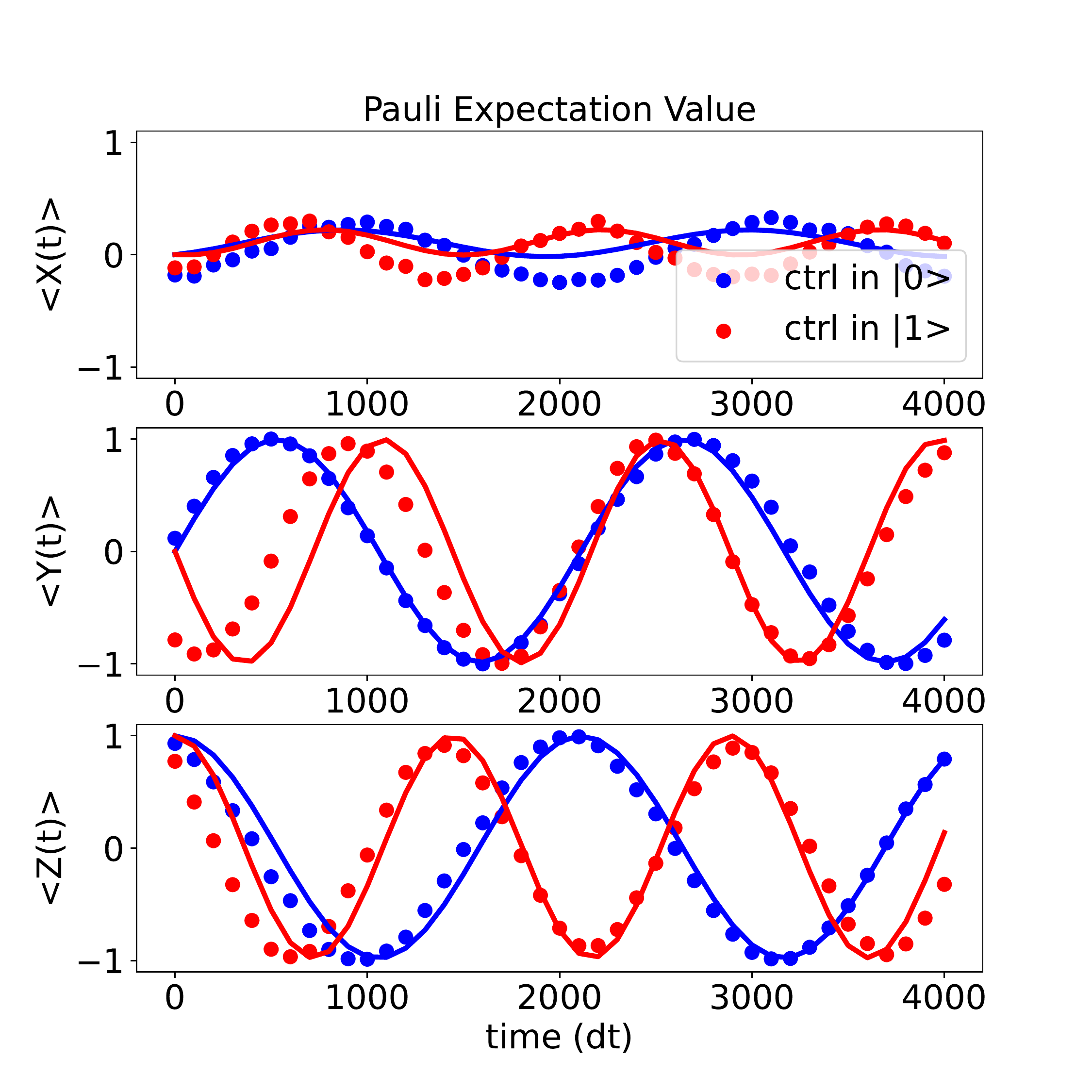}
    \caption{Calibration of the $CR$ pulse~\cite{aleksandrowicz2019qiskit}. The duration is swept to calculate the strength of different components. In gate-level quantum computing, $CR$ pulse usually needs calibration to become a $R_{ZX}(\pi/4)$ operation. In this process, phase shifts are added onto qubits to suppress the components like ZY. Other components are suppressed with echo $CR$ pulse and ShiftPhase operation. Extra ShiftPhase might also be needed to mitigate the ac-stark shift. In quantum pulse programs, such process is not required and $CR$ can be considered as an extra parameterized pulse, which enhances the flexibility of the pulse program.}
    \label{fig:HamTomo}
\end{figure}

\subsection{Fidelity estimation}

In quantum computing, it is important to calculate the fidelity of quantum states to determine if the implementation meets the desired accuracy. However, measuring fidelity can be computationally expensive, as it requires state tomography and comparing the state with noise-free simulation results. State tomography involves measuring quantum states in different basis to reconstruct the density matrix~\cite{cramer2010efficient}.
To mitigate this issue, one possible solution is to use the probability of successful trials (PST) as an approximation of the actual fidelity. This approach involves with a reversed program to restore the quantum states to their initial states. 
As is shown in Figure~\ref{fig:ZNE_demo}, a reversed circuit is appended to the original circuit. The PST of the measurement results indicate the level of noise in the quantum system.

\subsection{Randomized benchmarking in quantum computing}
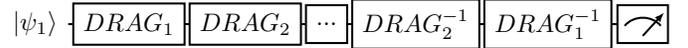
\begin{figure}[htbp]
  \centering
  \resizebox{0.5\textwidth}{!}
  {
    \begin{quantikz}[row sep=0.07cm,column sep=0.07cm]
    \lstick{$\ket{\psi_1}$} & \gate{DRAG_1} &\gate{DRAG_2} &\gate{...}  &\gate{DRAG_2^{-1}} &\gate{DRAG_1^{-1}} & \meter{} \\
    \end{quantikz}
  }
  \caption{Example of randomized benchmarking of parameterized pulses. Here we have a sequence of $DRAG$ pulses with random parameters and the reversed circuit. By measuring the proportion of measurement results that are the same as the initial state, we can obtain the fidelity estimation of the pulse sequence. The relationship between the estimated fidelities and the number of random pulses gives info about the property of the set of parameterized pulses. }
  \label{fig:quantum-circuit}
\end{figure}

Randomized benchmarking is a technique used in quantum computing to measure the average error rate of a set of quantum gates or operations~\cite{magesan2011scalable, mckay2019three, proctor2022measuring, gambetta2012characterization}. It is a statistical method that allows for the estimation of the error rate by performing a series of random circuits and comparing the output to the expected result.
% In randomized benchmarking, a sequence of random quantum gates is applied to a set of qubits, followed by an inverse sequence to return the qubits to their initial state. 
% The output of the circuit is compared to the expected output, and the probability of an error is estimated by repeating the process with multiple sequences of random gates.
By performing randomized benchmarking, we can obtain an estimate of the average error rate of a set of gates or operations, without the need for a complete characterization of each individual gate. 
This can help in the optimization of quantum hardware and algorithms by identifying and addressing error sources. 
To assess the fidelity of a gate sequence, one can compare noise-free simulation results with actual measurement results. 
However, as discussed earlier, the pulse simulator has limitations which make noise-free simulations unreliable, leading to inaccurate fidelity estimation of the pulse sequence. 
Therefore, we can enable randomized benchmarking for parameterized pulses when the pulse sequence is reversed and the probability of successful trials (PST) is used to approximate the fidelity of the pulse sequence, as illustrated in Figure~\ref{fig:quantum-circuit}.

\subsection{ZNE in quantum computing}

\begin{figure}
    \centering
    \includegraphics[width=0.5\textwidth]{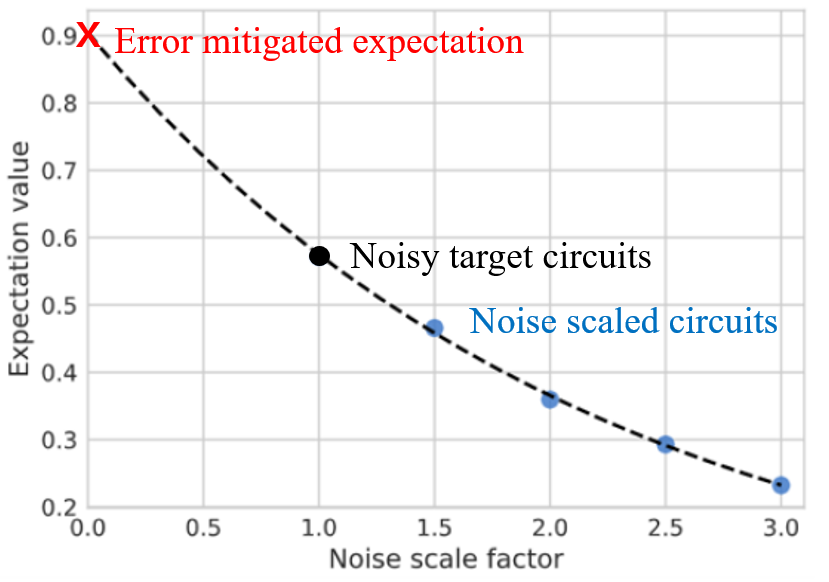}
    \caption{ZNE diagram. The noise-induced bias the target circuit can be mitigated by running noise scaled circuits and extrapolate the noisy expectations to zero noise scale which produces an error mitigated expectation value.}
    \label{ZNE diagram}
\end{figure}

To improve the NISQ-algorithm performance, zero-noise extrapolation (ZNE)~\cite{kandala2019error,giurgica2020digital} is proposed to mitigate the noise impact on observables expectation of output state. The basic idea of ZNE is inserting identity-equivalent gates to the target circuits, such as $G^\dagger$ and $G$. Ideally, these two gates will multiply to form identity operation. However, in experiment, due to the accumulation of noise, these redundant gates will introduce additional noise. The more noise involved, there is more bias on the observable expectation. Based on this observation, the bias versus noise scale decay curve can be fitted and extrapolate to the zero noise limit to get an error mitigated expectation value~\cite{li2017efficient, temme2017error, kandala2019error}. By applying ZNE, the noise-induced bias will be mitigated, at the cost of sample overhead in experiments due to the extrapolation step.
The implementation of ZNE contains two parts: noise scaling and extrapolation. The results of different noise level can be used to construct a noise model that characterizes the effect of noise on the observables of interest. This noise model is then used to extrapolate the results to the zero-noise limit, providing an estimate of the ideal outcome.
ZNE is particularly useful in situations where the noise level is high, and the computation is sensitive to small errors. It can be used to improve the accuracy of quantum algorithms by reducing the impact of noise on the computation. 

The key insight of ZNE is that the expecation value of measurement results change in a monotonic manner with the level of noise. This enables us to construct a simple model that captures the relationship between the expectation value and the noise level, by introducing additional controlled errors. With the model, we can deduce the expectation value of the so-called ``zero noise'' situation. To implement ZNE in gate-level quantum programs, a typical method is to reverse the program and append it to the original program. As is shown in the Figure~\ref{fig:ZNE_demo}, a reversed pulse schedule is added to the original circuit to restore the quantum states to their initial state.  Then, a copy of the original pulse schedule is appended to the end to achieve the same function as the original pulse schedule. This results in a quantum circuit with triple the noise, but identical function. We can obtain the expectation value of a quantum circuit with ``zero noise'' by comparing it to the expectation values of the same circuit with different noise levels.

% % Quantum Circuit

\section{Motivation}
%\tyl{this section has overlap with intro}

In this section, we will discuss the limitations of existing pulse-level works and pulse simulators, which have led to the unavailability of fidelity estimators and certain quantum error correction (QEC) techniques for pulse-level quantum programs.
% Recent pulse-level works lack consideration of realistic system Hamiltonians, as emphasized in SimuQ~\cite{peng2023simuq} and other approaches such as quantum optimal control (QOC) and pulse-level variational quantum algorithms (VQA) like Ctrl-VQE~\cite{} and Accqoc~\cite{}. 

The current pulse simulators, such as QuTiP~\cite{johansson2012qutip} and Qiskit pulse simulators~\cite{mckay2018qiskit}, suffer from inherent limitations. A major issue is the inconsistency between pulse simulation results and gate simulation results, indicating inconsistencies in the simulation process. Several factors contribute to these limitations, including a mismatch between the system Hamiltonian and pulse instruction parameters, as well as imprecision in solving ordinary differential equations (ODEs). Over time, inaccuracies in ODE solutions can accumulate, resulting in algorithmic errors that further widen the gap between ground truth and simulation results.
% In addition to the limitations discussed earlier, c
Current pulse simulators also suffer from large computation overhead. 
% The behavior of these simulators is similar to gate simulators but with many more gates, where each gate corresponds to a time slot of quantum pulses. Therefore, 
Even for a very short pulse circuit, the computation overhead can be comparable to simulating a large gate circuit. For instance, in Table \ref{tab:simu_time}, we demonstrate simulation time of a single Hadamard gate on pulse simulators with different numbers of qubits. There is a sharp increase in simulation time from a few seconds to a few minutes, when the qubit number is increased from five to seven, making pulse simulation expensive for pulse programs with more qubits. To simulate a pulse program with 100 pulses on a 7-qubit pulse simulator would take hours or even days to finish. 

% Drawbacks of current pulse-level work, dont' consider realistic system hamiltonian, (Yuxiang Peng's paper,,,,) (other QOC paper)(other pulse-level VQA paper)

% Drawbacks of pulse simulator, pulse simulation, (qutip, qiskit pulse simulator)(Give an example, show pulse simulation results differ from gate simulation results)

\begin{table}[]
    \centering
    \begin{tabular}{|c|c|c|}
    \hline
    \textbf{Backend Name} & \textbf{\#qubits} & \textbf{Simulation Time (s)} \\
    \hline
    FakeArmonk & 1 & 0.08 \\
    FakeAthens & 5 & 1.30 \\
    FakeQuito & 5 & 1.31 \\
    FakeBogota & 5 & 1.23 \\
    FakeLima & 5 & 1.28 \\
    FakeJakarta & 7 & 136.71 \\
    FakeCasablanca & 7 & 122.82 \\
    FakeLagos & 7 & 145.39 \\
    FakeNairobi & 7 & 123.90 \\
    \hline
    \end{tabular}
    \caption{The table shows the simulation time of a single Hadamard gate on pulse simulators with different numbers of qubits. There is a sharp increase in simulation time when the number of qubits changes from five to seven. This increase in simulation time makes it infeasible to simulate larger pulse programs with current pulse simulators.}
    \label{tab:simu_time}
\end{table}

% Need for profiling parameterized pulses

% Need for estimating pulse programs' fidelities

Estimating the fidelity of quantum states is critical for validating and characterizing the states generated by a quantum computer. 
However, this task is computationally expensive, and to accelerate the process of fidelity calculation, several techniques have been proposed. These techniques include variational quantum algorithms~\cite{cerezo2020variational}, machine learning-based approaches and statistical methods~\cite{yu2022statistical}. Additionally, the concept of ``classical shadow''~\cite{huang2020predicting, huang2022learning, huang2022provably} has been proposed for more efficient state tomography. 
However, in the scenario of quantum pulses, these techniques are not feasible due to the limitations of pulse simulators.

We propose to use reversed pulses to create a fidelity estimator for quantum pulses, which can be used to evaluate the noise level of pulse circuits and provide guidance for circuit design. By reversing the pulses, we can also conduct randomized benchmarking for parameterized pulses. For instance, if a pulse circuit design use parameterized $DRAG$ pulses with different amplitudes and angles, we can assess the noise performance of this set of $DRAG$ pulses with randomized benchmarking.
In future, if parameterized pulses of a certain shape is proposed, they can be evaluated with our randomized benchmarking technique to determine the error level.
% Although several error mitigation techniques have been developed for gate-level quantum programs, there has been relatively little research on designing error mitigation techniques specifically for pulse programs. While recent studies have employed pulse-level techniques such as dynamic decoupling for error mitigation, these techniques have not been specifically tailored for pulse-level programs.

We note that one of quantum pulses' promising applications is the variational quantum algorithm. Therefore, we implement a pulse-level VQE and adopt the ZNE method onto the task of calculating the lowest energy of a molecule. The ZNE method is integrated in the optimization loop, which means that the optimizer receives the ``zero noise'' expectation values and then update the parameters.

% Need for QEC on pulse programs

% \clearpage

\section{Techniques}
% In this section, we will dive into the technical details of pulse-level fidelity estimation and pulse-level zero noise extrapolation (ZNE). We will start by discussing the crucial step of reverse pulse scheduling, which is the key to fidelity estimator. We will provide a detailed analysis of the corresponding Hamiltonian and explain each step in the process while providing pseudocode for reference. Furthermore, we will explain how pulse-level VQE can be directly implemented and discuss how pulse-level ZNE can be applied to pulse-level VQE.
\subsection{Reversed pulse}
To reverse quantum pulses is a crucial step in our proposed technique. Reversing a pulse on a single qubit is a relatively straightforward process. By inserting a phase of $\pi$, the controlled part of the Hamiltonian can be inverted, effectively cancelling out the original pulse. As seen in the Hamiltonian of the qubit system, 
$$
    H(t)/ \amshbar = - \frac{\omega_q}{2}\sigma_z -  \frac{\Omega(t)}{2}\sigma_x
$$
the first term represents the rotation of the qubit around the Z-axis, and the second term represents the effect of pulses on the DriveChannel. The addition of a phase of $\pi$ to the pulse results in the cancellation of the second term. The first term does not have any effect on the pulse's functionality, as the reference frame is also rotating around the Z-axis with the same frequency.

\lstset{
  language=Python,
  basicstyle=\ttfamily\small,
  keywordstyle=\color{blue},
  % numbers=left,
  frame=single,
  backgroundcolor=\color{gray!20},
  breaklines=true
}

\begin{figure*}
\centering
\begin{pythoncode}
def rev_sched(orig_sched):
    sched = Schedule()
    duration = orig_sched.duration
    for inst in reversed(orig_sched.instructions):
        if(isinstance(inst[1],qiskit.pulse.instructions.phase.ShiftPhase)):
            rev_inst = ShiftPhase(-inst[1].phase,inst[1].channel) 
            sched = sched.insert(duration-inst[0],rev_inst)
        elif(isinstance(inst[1],qiskit.pulse.instructions.play.Play)):
            rev_inst = copy.deepcopy(inst[1])
            rev_inst.pulse._params['amp'] = inst[1].pulse.amp
            rev_inst.pulse._params['angle'] = inst[1].pulse.angle + np.pi
            sched = sched.insert(duration-inst[0]-rev_inst.duration,rev_inst)
    return sched
\end{pythoncode}
\caption{Python codes that reverse a pulse program. If a ShiftPhase operation is encountered, a new ShiftPhase with opposite phase will be inserted into the schedule. If a parameterized pulse is detected, a new parameterized pulse with additional phase $\pi$ will be inserted into the schedule in the corresponding position.}
\label{python_code}

\end{figure*}

In Qiskit pulse programming, the ShiftPhase operation is used to manipulate the phase of a pulse on a particular channel for a single qubit. To undo the effect of the ShiftPhase operation, we simply need to apply the phase change in the opposite direction.
To reverse a multi-qubit pulse is more complicated. We choose $CR$ pulses as our target pulse to reverse, since it can produce entanglement. When $CR$ pulses are added on the ControlChannel of qubit, the Hamiltonain of the qubit system is :
\begin{align*}
\tilde{H}_{\rm eff}^{\rm CR} = - \frac{\Delta_{12}}{2}\sigma_1^z + \frac{\Omega(t)}{2} \left(\sigma_2^x - \frac{J}{2\Delta_{12}} \sigma_1^z \sigma_2^x \right)
\end{align*}
The equation contains three terms: the rotation around the Z-axis, the rotation of the second qubit around the X-axis denoted by IX, and the ZX interaction term that creates entanglement. 
% We can generate maximally entangled state by placing qubit 0 in an equal superposition of state 0 and state 1. Then, by applying the cross resonance gate with a duration that corresponds to a $\pi/4$ rotation around the X-axis, the entangled state is produced.
To cancel out the last two terms, we need to apply a phase shift of $\pi$ on the phase of the $CR$ pulse that acts on the control channel.
The $CR$ pulse is driving the control qubit at the frequency of the target qubit, and there is a coupler in between two qubits. The state of the control qubit determines the frequency of Rabi oscillations of the target qubit. Therefore, after a certain time, the $CR$ pulse can achieve a controlled $\pi$ phase shift.
When the pulse is reversed, the phase difference will change with time in the other direction, which then reverts the quantum state.

\subsection{Pulse-level fidelity estimator}

\begin{figure}
    \centering
    \includegraphics[width=0.5\textwidth]{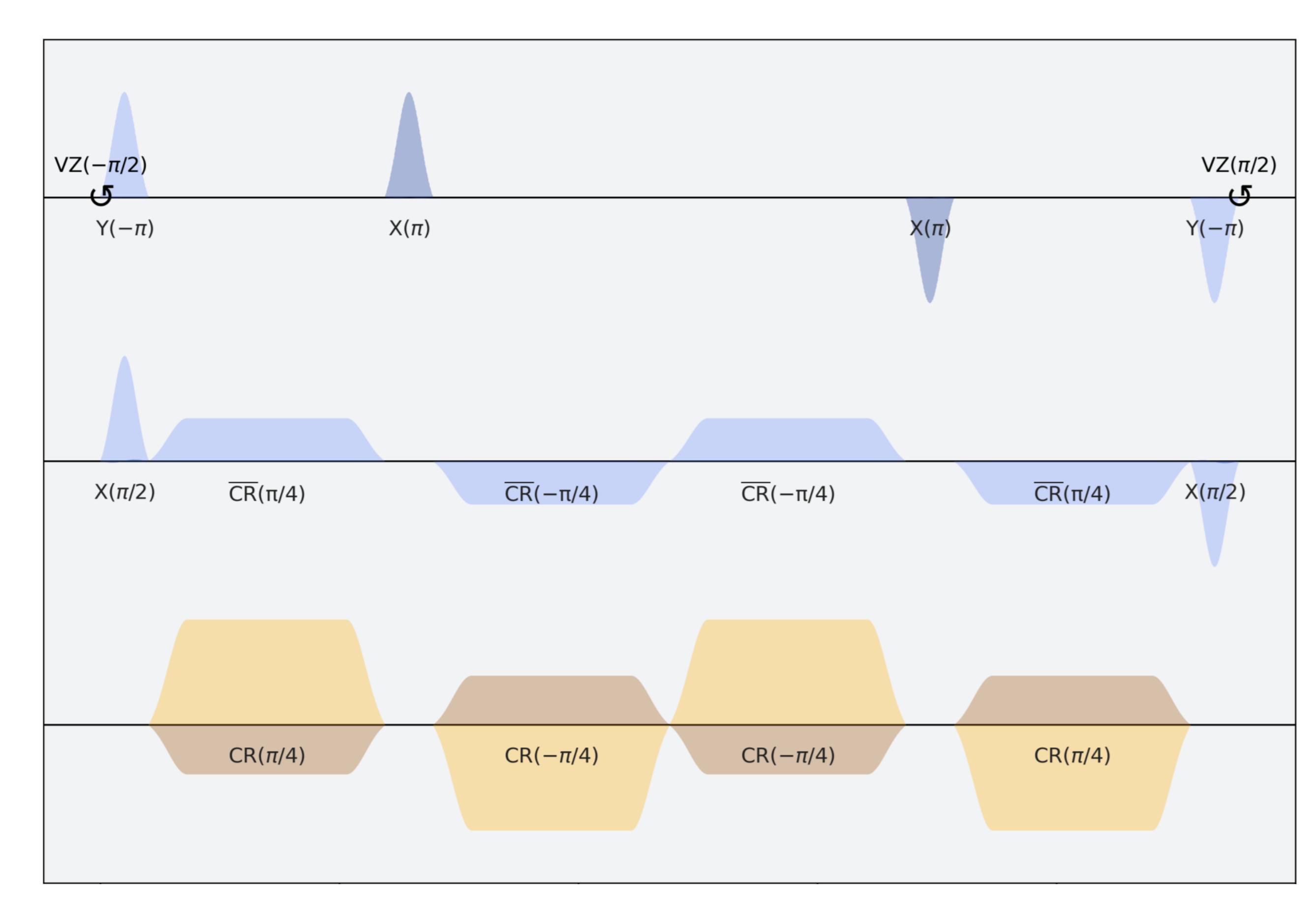}
    \caption{Example of original pulse schedule with its reversed pulse schedule. We can see that the pulses have phase difference of $\pi$ and the order is also reversed. }
    \label{fig:CXCXrev}
\end{figure}

In pulse programs, to make sure that reversed pulses are able to cancel the original pulses in the correct order, we need to record the timestamp of each pulse and insert the reversed pulse accordingly. As shown in Figure~\ref{python_code}, the ShiftPhase operation is directly reversed with negative phase, other pulses are inserted to the position that maintains the structure of the pulse program.
An example is presented in the Figure~\ref{fig:CXCXrev}, the reversed CX pulse shows symmetric structure as the original CX pulse.
After we obtain the reversed pulse program. it is appended to the end of the original pulse program.
It has been demonstrated in recent work~\cite{wang2022quest} that the PST of appended circuit is a good proxy for the fidelity.

\subsection{Pulse-level VQE}

\begin{figure*}[htb]
    \centering
    \includegraphics[width=0.88\textwidth]{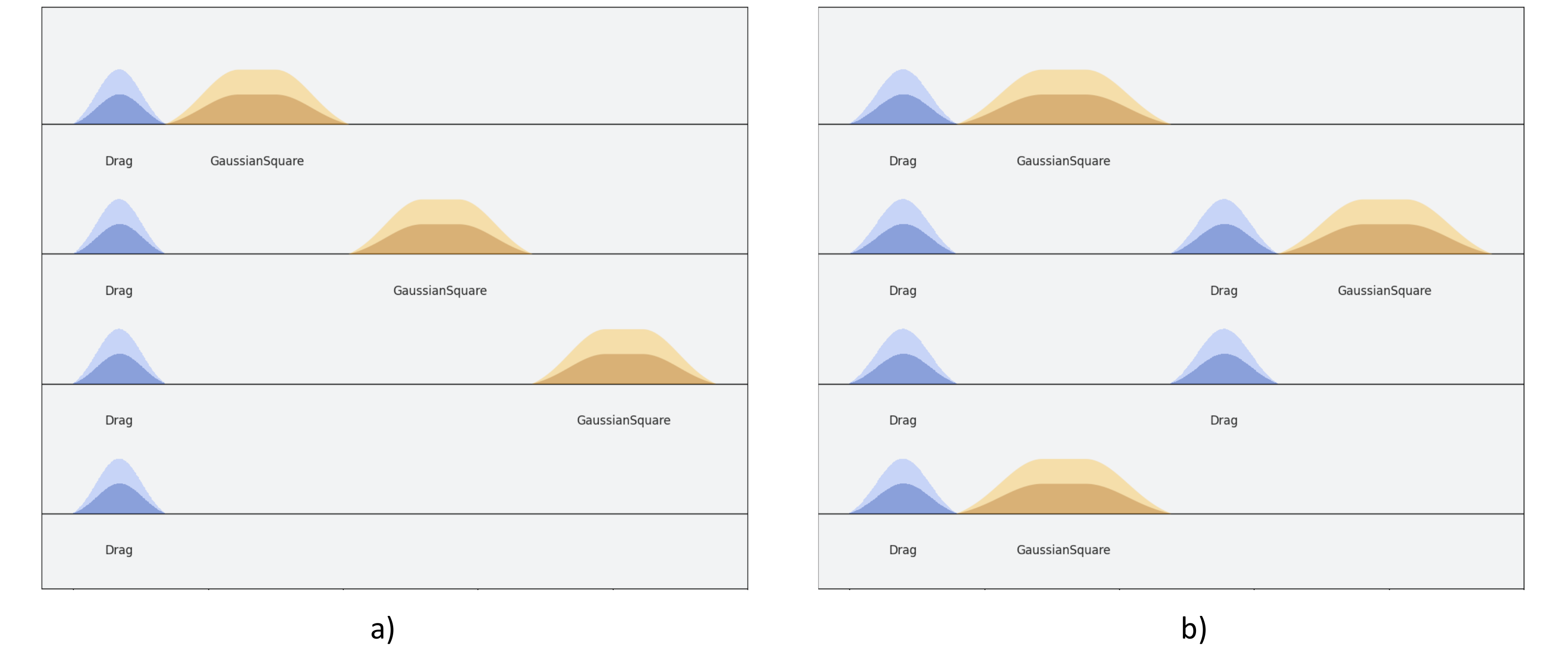}
    \caption{Pulse schedules of ansatz used in VQE, both are hardware-efficient ansatz inspired. In the training process, the parameters of these pulses are updated to minimize the expecation value. The $GaussianSquare$ pulses in the figure are applied onto the ControlChannel to introduce entanglement between qubits. The $DRAG$ pulses are applied on the DriveChannels.}
    \label{ansatz}
\end{figure*}

Recent works~\cite{egger2023study, meitei2021gate, liang2022pan, liang2023towards} have pointed out that parameterized pulses can improve the performance of the Variational Quantum Eigensolver (VQE) by eliminating the need for redundant gate-level compilation. The single-qubit pulse and two-qubit pulse are the fundamental building blocks of pulse programs for VQE.
For the single-qubit pulse, the $DRAG$ pulse shape provides the best fidelity and lowest energy leakage, making it the ideal choice. The $CR$ gate is the simplest pulse that can trigger entanglement between two qubits, making it the preferred option for implementing two-qubit pulses. The two-qubit interaction is achieved by applying pulses at the frequency of the second qubit onto the first qubit.
The single-qubit and two-qubit pulses share the same set of parameters, including amplitude, duration, angle, shape, and frequency. With $DRAG$ and $CR$ as building blocks, we can construct pulse circuits for VQE similarily to the construction of the hardware-efficient ansatz (HEA). 
In the HEA-inspired pulse ansatz, $DRAG$ pulses are inserted on each single qubit and $CR$ pulses are applied to all available qubit connections.
Then the optimizer iteratively updates the parameters of the pulses based on the final expectation value obtained through measurements. 
To be more specific, the Hamiltonian of the target system is usually expressed as a weighted sum of Pauli strings. 
The expectation value is then computed as the weighted sum of the expectation values obtained from different pulse circuits corresponding to each Pauli string.

\subsection{ZNE for pulse-level VQE}

To enable ZNE for pulse-level VQE tasks, we need to construct the VQE circuit with triple noise and compute the loss function with the ZNE method in each training iteration. One straightforward approach is to utilize a linear model to determine the expectation value under ``zero noise'' conditions. For instance, when calculating the energy of a H2 molecule, the original ansatz circuit produces an expectation value of -1.8518 H, whereas the triple noise ansatz circuit returns an expectation value of -1.8464 H. Therefore, the noise introduces a bias of +0.0054H. Assuming that the expectation value is linear with the noise level, we can then obtain the ``zero noise'' expectation value of -1.8545 H, which is closer to the ideal value of -1.8581 H. This ZNE technique reduces the deviation from the ideal value by 50\%, which is a significant improvement in quantum chemistry applications.

% For our experiment, we consider the molecules of $H_2\ and\ LiH$. 
% They contain two electrons. 
% Our VQE aims to calculate their ground energies at bonding lenth configuration. 
% We adopt the minimum sets of Slater-type-orbitals-Guassian (STO-3G) functions as basis to expanding the electronic Hamiltonian. 
% Under this configuration the ansatz can capture the interactions of two spin-orbitals at the most inner shell with two qubits. The circuits start with Hartree-Fock state $|HF\rangle$, and the hardware efficient anstaz is applied to the HF state to capture interactions out of the HF state by exciting electrons from ground orbitals. By minimizing the energy of the prepared state and and optimizing the ansatz parameters the ground state can be approach from which we cna measure the ground energy.

% 1. how to reverse pulse program
% 2. theory to support the reverse of pulse program
% 3. source of errors of pulse program
% 4. theory support for mirror benchmarking (why orig + rev can estimate true fidelity)
% 5. 

% \clearpage
% Another page of techniques
% \clearpage

\section{Results}

\subsection{Experimental setup}
% Our experiments are conducted both on simulators and NISQ machines.
We evaluate the performance of fidelity estimator on IBM quantum systems: $ibmq\_montreal,\  ibm\_auckland$ and $ibm\_geneva$. 
We evaluate the performance of pulse-level ZNE on pulse simulators which are based on Qiskit-Aer.
% Simulators are based on Qiskit-Aer and used to evaluate the performance of pulse-level ZNE.
Simulations are run on a server with two Intel Xeon E5-2630 CPUs, 64 GB DRAM, with CentOS 7.4 as the operating system.
% And we choose VQE problems as our evaluation tasks, which consist of several molecules including $H_2$, and $HeH+$.
In VQE tasks, we use hardware-efficient-ansatz inspired ansatz \cite{liang2023towards}. Examples are shown in Figure \ref{ansatz}. We use COBYLA as optimizer and set the number of shots of quantum program to 1024.

\subsection{reversed pulse and fidelity estimator}

\begin{figure}
    \centering
    \includegraphics[width=0.5\textwidth]{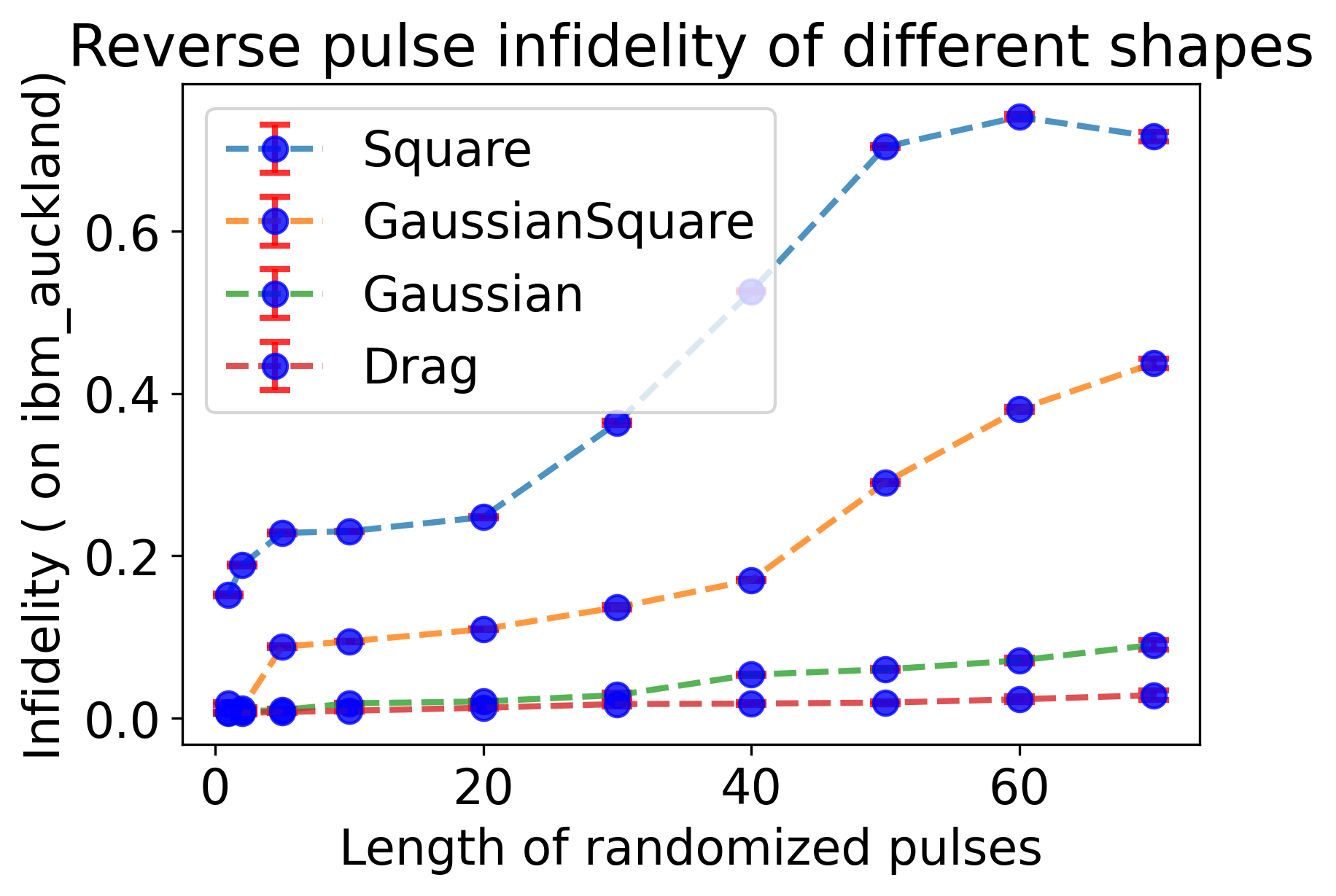}
    \caption{We demonstrate the reverse pulse fidelity of different pulse shapes. The figure suggests that the $Square$ pulse exhibits the most significant error rate, while Gaussian and $DRAG$ pulses have comparatively lower errors. We ensure that each pulse have the same duration. 
    % This might explain why the $GaussianSquare$ pulse also has high infidelity. 
    }
    \label{fig:output}
    
\end{figure}

\begin{figure}[htb]
    \centering
    \includegraphics[width=0.5\textwidth]{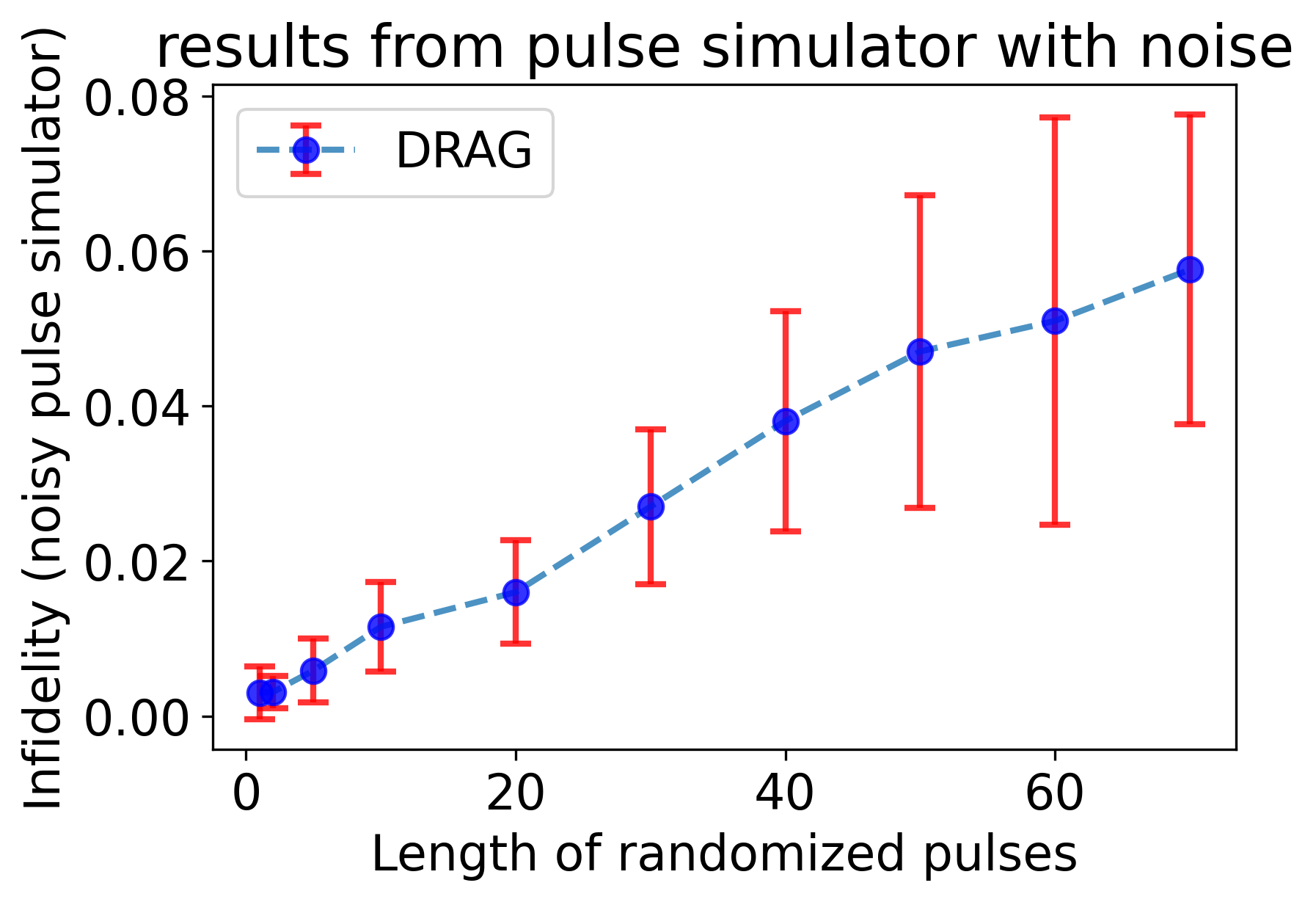}
    \caption{The curve represents the relationship between the infidelity and the number of pulses with randomized parameters. The results are obtained from a modified pulse simulator. The noise is inserted into the pulse simulator by adding Gaussian noise to pulses' amplitudes and white noise to the measurement outputs. The parameters of the noise are tuned to obtain a curve that is similar with the results from real quantum computers.}
    \label{pulsesimulatornoiseresults}
    
\end{figure}

\begin{figure}[htb]

    \centering
    \includegraphics[width=0.5\textwidth]{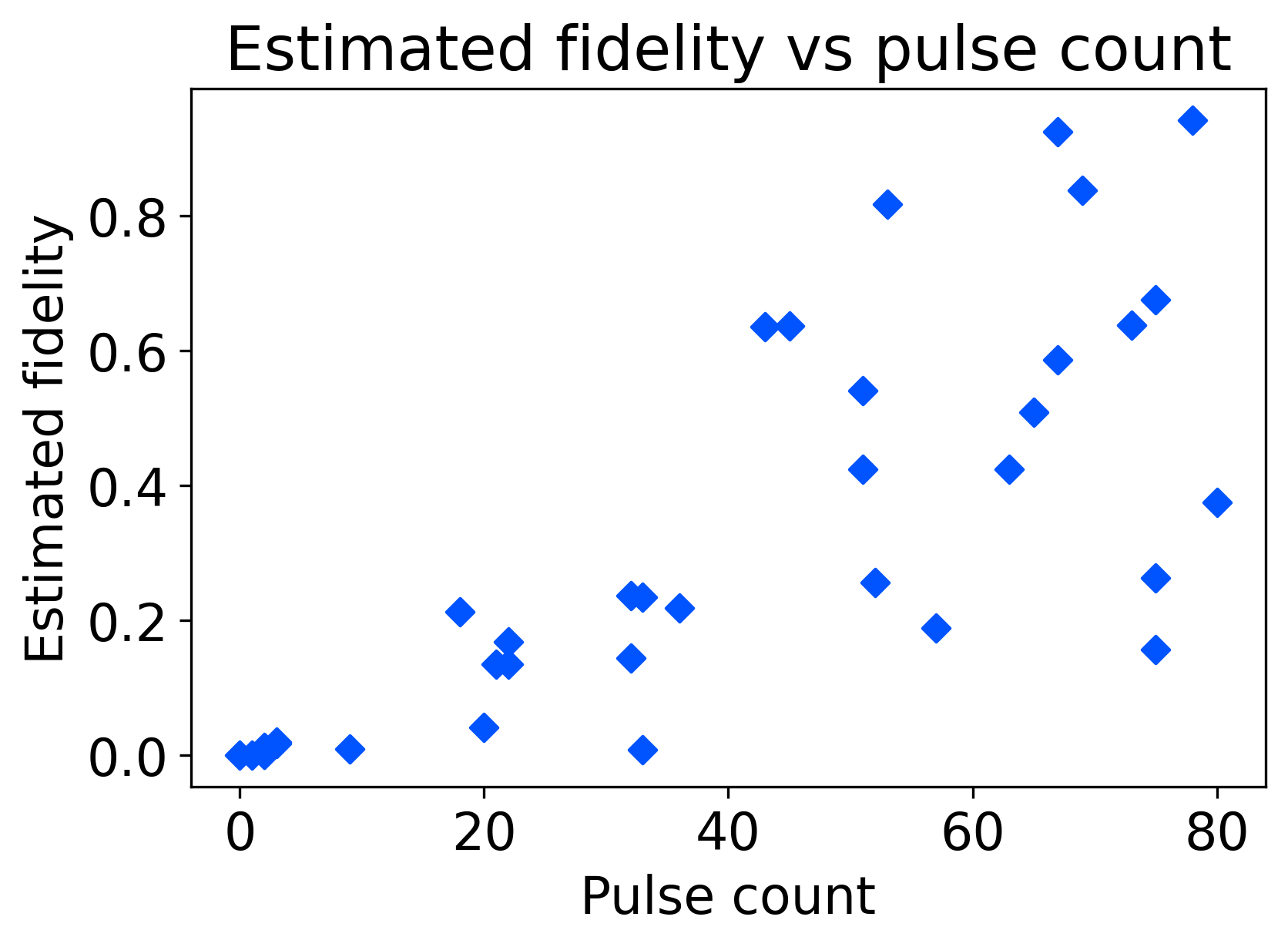}
    \caption{The estimated fidelity with the number of pulses in the pulse schedule. The data is collected from a group of randomly generated circuits. }
    \label{estimatedfidelityvspulsecount}
    
\end{figure}

We test the reverse of pulses on IBM's quantum computers and observe an accuracy of over 99.999\% for the reversed operation for $DRAG$ pulses, which validates our method to reverse the pulse.
$DRAG$ pulses have great properties such that the leakage and the phase errors are suppressed. We also demonstrated other shapes of pulses on quantum computers and observe different levels of inaccuracy in the Figure~\ref{fig:output}. 
We can tell from the figure that $DRAG$ pulse has the lowest infidelity. 
The infidelity of Guassian pulse is higher, which may be caused by the accumulated phase error. 
The infidelity of both the $Square$ pulse and $GaussianSquare$ pulse is considerably higher. This elevated infidelity comes from the unwanted interaction between the qubit and the environment, which can be caused by the high-frequency component of the pulse. The results show that decoherence error is not the only factor that contributes to the noise.

Since we have limited access to IBM's quantum devices, we employ pulse simulators for certain experiments. To ensure that the simulated results resemble closely with those obtained from actual quantum machines, we manually introduce noise into the pulse simulator. The insertion of noise is achieved by adding a Gaussian noise to the amplitudes of pulses and a white noise to the measurement results. The curve in Figure~\ref{pulsesimulatornoiseresults} illustrates how we replicate the curve of $DRAG$ pulses' infidelity observed in real quantum computers. We can see that the trend of the curve and error bars close ensemble the results from real quantum computers in Figure~\ref{montreal_RB_RES}. We use this pulse simulator for fidelity estimation of pulse programs and ZNE experiment for pulse-level VQE.
Due to the mismatch of pulse parameters and the underlying system Hamiltonian, we are unable to produce the same results from gate simulator and pulse simulator. To be more specific, if a gate circuit is transformed into a pulse circuit and executed on pulse simulator, the outcomes do not match with the results from gate simulator. Therefore, we cannot find a reasonable reference to validate our estimated fidelity. In Figure~\ref{estimatedfidelityvspulsecount}, we present the results of our fidelity estimator. The curve reveals a positive correlation between the infidelity and the number of pulses within the circuit, which aligns with our expectations.

\subsection{Randomized benchmarking for parameterized pulses}

\begin{figure*}[]
    \centering
    \includegraphics[width=0.9\textwidth]{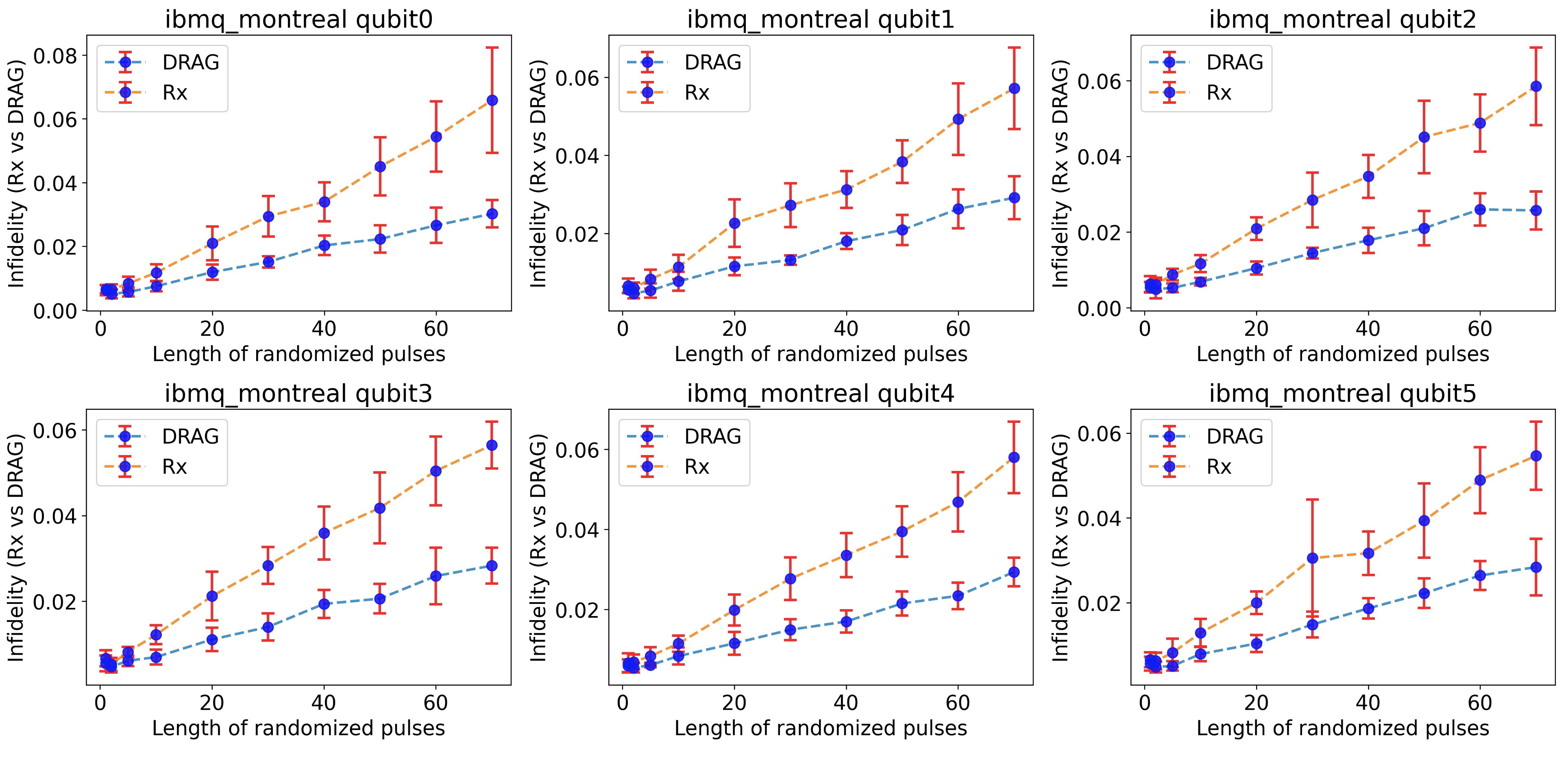}
    \caption{Characterizing the performance of randomized $R_x$ pulses and $DRAG$ pulses on qubit 0 to qubit 5 of ibmq\_montreal. In IBM's quantum devices, the $R_x$ gate is decomposed into two $DRAG$ pulses with a ShiftPhase operation in between them. The ShiftPhase operation accurately determines the rotation angles. This operation is executed through a phase change in the control signals, not inducing any latency. Therefore, the duration of the $R_x$ gate is twice that of its $DRAG$ components. The duration correlation is also reflected in the curves, where the infidelity of the $R_x$ gate is approximately double that of the $DRAG$ pulse.
    }
    \label{montreal_RB_RES}
\end{figure*}

\begin{figure*}[h!]
    \centering
    \includegraphics[width=0.9\textwidth]{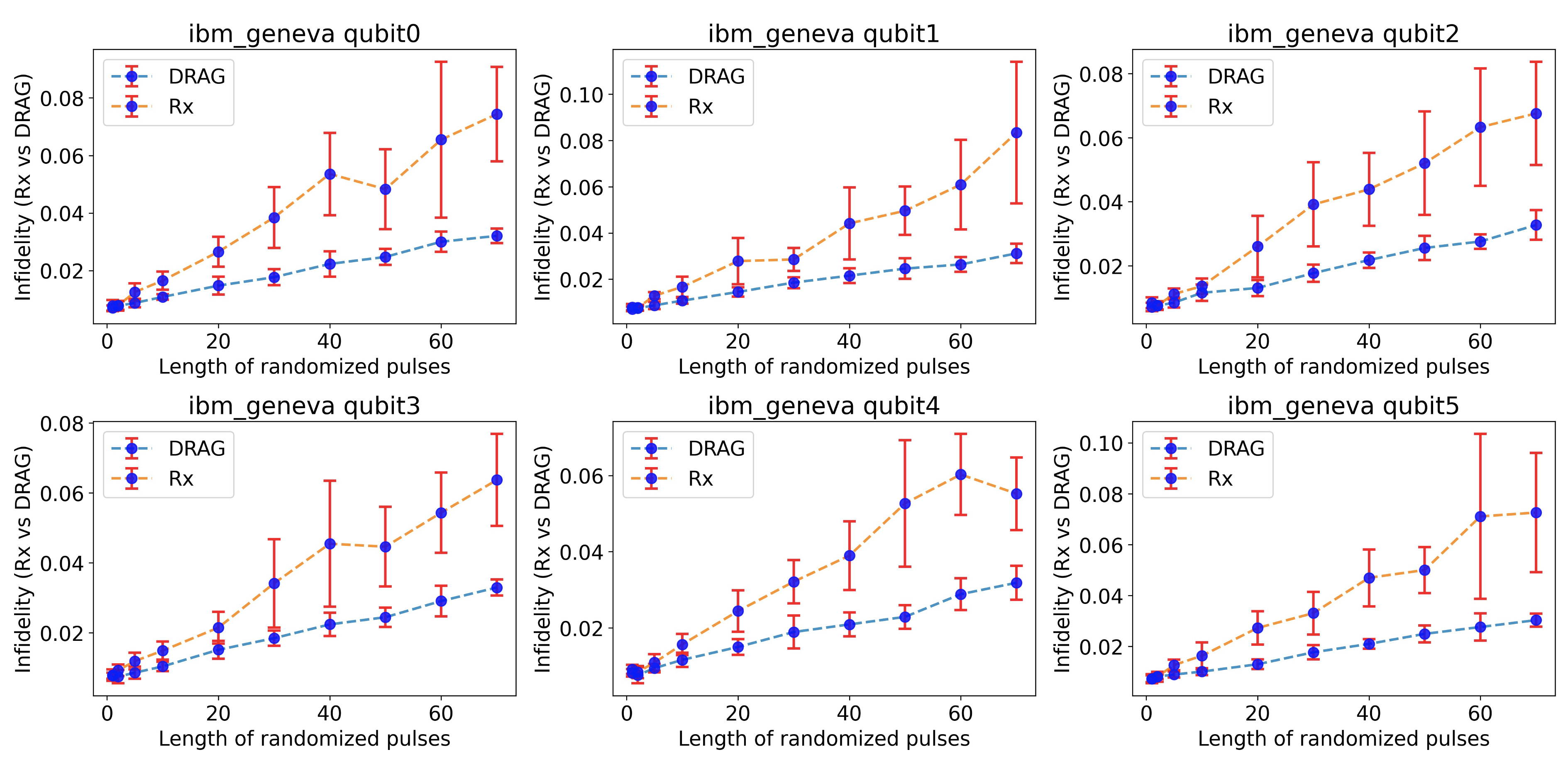}
    \caption{Characterizing the performance of randomized $R_x$ pulses and $DRAG$ pulses on qubit 0 to qubit 5 of ibm\_geneva. Similar to the results for ibmq\_montreal, the infidelity per $R_x$ gate is almost twice the infidelity per $DRAG$ pulse. We also observe a relatively linear correlation between the infidelity and the number of randomized pulses. Since the infidelity is calculated through counting the probability of successful trials, such results align with the key insight of ZNE, where the expecation values change monotonically with the level of noise.}
    \label{geneva_rb_res}
\end{figure*}

We perform randomized benchmarking on different qubits across multiple quantum devices to analyze parameterized pulses. Specifically, we assign random amplitudes to $DRAG$ pulses and create a pulse sequence. We then reverse the pulse sequence and collect the results. As shown in the Figure~\ref{montreal_RB_RES} and the Figure~\ref{geneva_rb_res}, we note that the infidelities exhibit nearly linear growth as the number of randomized pulses increases for qubits 0 to qubit 5 on the $ibmq\_montreal$ and $ibm\_geneva $ machines.
Also, we compare the infidelity of the $R_x$ gates with that of the $DRAG$ pulses. On IBM's quantum computers, the $R_x$ gates are decomposed into three components: two $DRAG$ pulses performing $R_x(\pi/2)$ and one ShiftPhase operation to achieve the desired rotation. This decomposition method allows us to obtain an accurate rotation angle for the $R_x$ gates. We note that the ShiftPhase operation is achieved without actual pulse or operation on quantum bits. ShiftPhase is only changing the phase of the control signals, which results in no increase of the circuit latency. So the latency of the $R_x$ gate is double the latency of its $DRAG$ components, which is also reflected in the curves. We can see in Figure~\ref{montreal_RB_RES} that the infidelity of $R_x$ gate is roughly double the infidelity of $DRAG$ pulse. On the other hand, $DRAG$ pulses can produce the same rotation with less latency, but we lose track of the precise rotation angle. The curves in the figures illustrate that $R_x$ pulses exhibit higher infidelity compared to $DRAG$ pulses. 
For general quantum programs, inaccurate rotation angles are not favorable. But in variational quantum algorithms where parameters are constantly updated, we may prioritize ``less circuit latency'' over ``accurate rotation angles''. This is because that control errors such as under-rotation and over-rotation can be effectively corrected by tuning the parameters of the pulses in pulse-level variational quantum algorithms. 
In Figure~\ref{fig:output}, we can see the results of randomized benchmarking for pulses of different shapes. Among these shapes, the $DRAG$ pulse produces the lowest infidelity for 0.04\% while the $Square$ pulse produces the highest infidelity of 1\%. Through randomized benchmarking for quantum pulses, we can distinguish the ``good'' pulses from ``bad'' pulses, which is helpful in the design of pulse programs. For example, if a new optimal control algorithm is developed and generates a sequence of pulses. The pulse-level randomized benchmarking can help evaluate the performance of the pulses.

\begin{figure}[htb]
    \centering
    \includegraphics[width=0.5\textwidth]{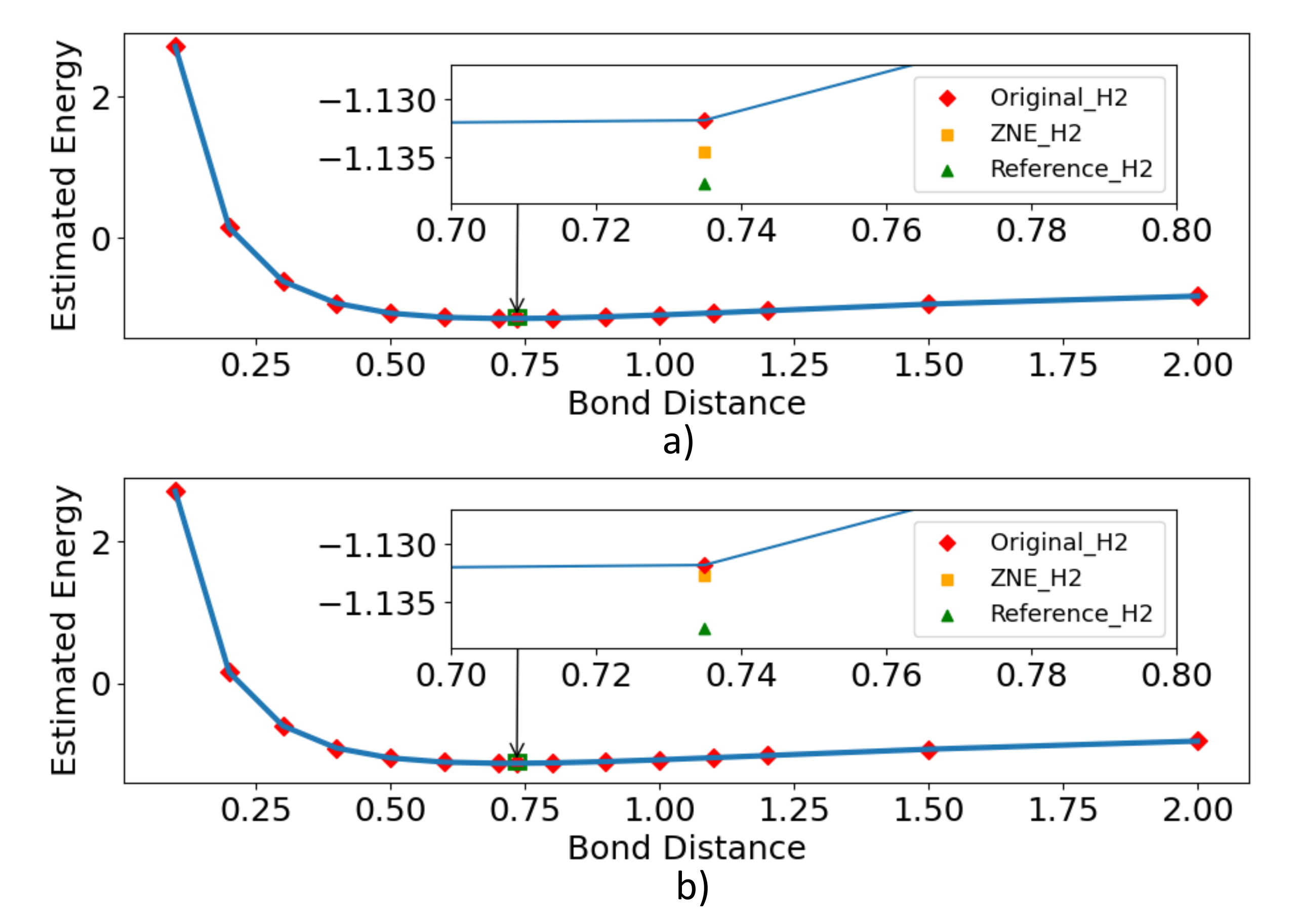}
    \caption{Energy-distance curve of H2 molecule. The values of the ground-state energy is obtained with pulse-level VQE. The Original\_H2 and the curve represent the pulse-level VQE results without ZNE. The ZNE\_H2 represents the value of the ground-state energy with ZNE. We also mark the ideal value in the curve for reference. The two figures are obtained from two different ansatz. Results show that the ZNE method works for both ansatz designs.}
    \label{vqe}
\end{figure}

\subsection{Pulse-level VQE and ZNE}

\begin{table}[]
\centering
\begin{tabular}{|p{2cm}|p{1cm}|p{1cm}|p{1cm}|p{1.5cm}|}
\hline
Model                   & Molecule      & Dist           & Accuracy         & Deviation \\ \hline
Ansatz 1                & H2            & 0.735          & 99.52\%          & 0.00547 H            \\ \hline
\textbf{Ansatz 1 + ZNE} & \textbf{H2}   & \textbf{0.735} & \textbf{99.76\%} & \textbf{0.00272 H}   \\ \hline
Ansatz 2                & H2            & 0.735          & 99.52\%          & 0.00547 H            \\ \hline
\textbf{Ansatz 2 + ZNE} & \textbf{H2}   & \textbf{0.735} & \textbf{99.60\%} & \textbf{0.00458 H}   \\ \hline
Ansatz 1                & HeH+          & 1.0            & 99.71\%          & 0.00537 H            \\ \hline
\textbf{Ansatz 1 + ZNE} & \textbf{HeH+} & \textbf{1.0}   & \textbf{99.93\%} & \textbf{0.00206 H}   \\ \hline
Ansatz 2                & HeH+          & 1.0            & 99.71\%          & 0.00537 H            \\ \hline
\textbf{Ansatz 2 + ZNE} & \textbf{HeH+} & \textbf{1.0}   & \textbf{99.98\%} & \textbf{0.00064 H}   \\ \hline
\end{tabular}
\caption{Comparison between the original VQE results and the results after ZNE is applied. An average reduction of 54.1\% is achieved in the deviation from the ideal energy values.}
\label{result}
\end{table}

Since we can reliably reverse a quantum pulse program, we have use this technique in the pulse-level Variational Quantum Eigensolver (VQE) to implement the Zero Noise Extrapolation (ZNE) method. To increase the noise level of the original VQE pulse program, we manually triple it by appending a reversed program and an original program. This results in a pulse program that maintains the same functionality but has triple the noise.
Once we obtain the expectation values from both the original circuit and the triple-noise circuit, we assume that the expectation value changes linearly with the level of noise. With this assumption, we can deduce the ``zero noise'' expectation value. We then perform the optimization with this ``zero noise'' expectation value. This approach enables us to obtain final energy values that are closer to the ideal values, which is important in quantum chemistry applications.
To validate our methods, we conducted tests with different ansatz and molecules. In all of our experiments, the results consistently demonstrated a reduction in deviation from the ideal value. This improvement is shown in the Table~\ref{result}. Our approach produces an average reduction of 54.1\% in the deviation from the ideal energy value.
We select two energy-distance curves to demonstrate that our ZNE results are closer to the reference values. The curves also prove that pulse-level VQE is a valid method for quantum chemistry problems. The parameterized single-qubit and two-qubit gates can effectively explore the state space while maintaining the trainability.

\section{Conclusion}

In this paper, we use reversed pulse programs to create a fidelity estimator and to implement Zero Noise Extrapolation (ZNE) on pulse-level quantum programs. 
The reversed pulse program is achieved by inverting the pulse schedule, keeping the same amplitude for each pulse but introducing an additional phase of $\pi$. 
With the reversed pulse program, we develop a fidelity estimator which results in a final state identical to the initial state under noise-free conditions. So we use the Probability of Successful Trials (PST) as our measure of noise level. 
With this fidelity estimator, we perform randomized benchmarking on parameterized pulses of various shapes and $R_x$ gates on IBM's quantum devices.
Our results indicate that the $DRAG$ pulse offers superior fidelity, which is approximately 0.04\% infidelity per $DRAG$ pulse of 160 $dt$ or 35.5 $ns$. 
The infidelity per $R_x$ gate is twice as much as the infidelity per $DRAG$ pulse.
However, this advantage is brought by the $DRAG$ pulse's incapacity to accurately control the rotation angle. 
Due to limited access to quantum computers, a part of our experiments are conducted on pulse simulators. 
We modify the pulse simulator by adding Gaussian noise to pulse amplitudes and white noise to the measurement. 
Then we adjust the noise parameters to align the behaviour of the modified simulator with that of real quantum computers.
With the modified pulse simulator, we estimate the fidelity for a set of quantum programs. 
The results suggest a positive correlation between the circuit's pulse count and infidelity. 
Then we execute a pulse-level Variational Quantum Eigensolver (VQE) program, in which the ansatz is inspired by HEA and is composed of $DRAG$ and $CR$ pulses. 
The VQE tasks include estimating the ground state energy of $H_2$ and $HeH^+$ at a specific bond distance. 
We implement ZNE on VQE tasks with the reversed pulse program, resulting in an average deviation reduction of 54.1\% from the optimal value. 
In conclusion, our work demonstrates that reversed pulses can be effectively used for pulse-level randomized benchmarking, fidelity estimation, and zero noise extrapolation.

% \FloatBarrier % Prevents floats from moving beyond this point

% \clearpage
%%%%%%% -- PAPER CONTENT ENDS -- %%%%%%%%

%%%%%%%%% -- BIB STYLE AND FILE -- %%%%%%%%
\bibliographystyle{IEEEtranS}
\bibliography{refs}

% Generated by IEEEtranS.bst, version: 1.13 (2008/09/30)
\begin{thebibliography}{10}
\providecommand{\url}[1]{#1}
\csname url@samestyle\endcsname
\providecommand{\newblock}{\relax}
\providecommand{\bibinfo}[2]{#2}
\providecommand{\BIBentrySTDinterwordspacing}{\spaceskip=0pt\relax}
\providecommand{\BIBentryALTinterwordstretchfactor}{4}
\providecommand{\BIBentryALTinterwordspacing}{\spaceskip=\fontdimen2\font plus
\BIBentryALTinterwordstretchfactor\fontdimen3\font minus
  \fontdimen4\font\relax}
\providecommand{\BIBforeignlanguage}[2]{{%
\expandafter\ifx\csname l@#1\endcsname\relax
\typeout{** WARNING: IEEEtranS.bst: No hyphenation pattern has been}%
\typeout{** loaded for the language `#1'. Using the pattern for}%
\typeout{** the default language instead.}%
\else
\language=\csname l@#1\endcsname
\fi
#2}}
\providecommand{\BIBdecl}{\relax}
\BIBdecl

\bibitem{aleksandrowicz2019qiskit}
G.~Aleksandrowicz, T.~Alexander, P.~Barkoutsos, L.~Bello, Y.~Ben-Haim,
  D.~Bucher, F.~J. Cabrera-Hernández, J.~Carballo-Franquis, A.~Chen, C.-F.
  Chen, J.~M. Chow, A.~D. Córcoles-Gonzales, A.~J. Cross, A.~Cross, o.~bw,
  J.~Cruz-Benito, C.~Culver, S.~D. L.~P. González, E.~D.~L. Torre, D.~Ding,
  E.~Dumitrescu, I.~Duran, P.~Eendebak, M.~Everitt, I.~F. Sertage, A.~Frisch,
  A.~Fuhrer, J.~Gambetta, B.~G. Gago, J.~Gomez-Mosquera, D.~Greenberg,
  I.~Hamamura, V.~Havlicek, J.~Hellmers, Å.~Herok, H.~Horii, S.~Hu,
  T.~Imamichi, T.~Itoko, A.~Javadi-Abhari, N.~Kanazawa, A.~Karazeev,
  K.~Krsulich, P.~Liu, Y.~Luh, Y.~Maeng, M.~Marques, F.~J. Martín-Fernández,
  D.~T. McClure, D.~McKay, S.~Meesala, A.~Mezzacapo, N.~Moll, D.~M. Rodríguez,
  G.~Nannicini, P.~Nation, P.~Ollitrault, L.~J. O'Riordan, H.~Paik, J.~Pérez,
  A.~Phan, M.~Pistoia, V.~Prutyanov, M.~Reuter, J.~Rice, A.~R. Davila, R.~H.~P.
  Rudy, M.~Ryu, N.~Sathaye, C.~Schnabel, E.~Schoute, K.~Setia, Y.~Shi,
  A.~Silva, Y.~Siraichi, S.~Sivarajah, J.~A. Smolin, M.~Soeken, H.~Takahashi,
  I.~Tavernelli, C.~Taylor, P.~Taylour, K.~Trabing, M.~Treinish, W.~Turner,
  D.~Vogt-Lee, C.~Vuillot, J.~A. Wildstrom, J.~Wilson, E.~Winston, C.~Wood,
  S.~Wood, S.~Wörner, I.~Y. Akhalwaya, and C.~Zoufal, ``Qiskit: An open-source
  framework for quantum computing,'' 2019.

\bibitem{alexander2020qiskit}
T.~Alexander, N.~Kanazawa, D.~J. Egger, L.~Capelluto, C.~J. Wood,
  A.~Javadi-Abhari, and D.~C. McKay, ``Qiskit pulse: programming quantum
  computers through the cloud with pulses,'' \emph{Quantum Science and
  Technology}, vol.~5, no.~4, p. 044006, 2020.

\bibitem{barron2020measurement}
G.~S. Barron and C.~J. Wood, ``Measurement error mitigation for variational
  quantum algorithms,'' \emph{arXiv preprint arXiv:2010.08520}, 2020.

\bibitem{blatt2012quantum}
R.~Blatt and C.~F. Roos, ``Quantum simulations with trapped ions,''
  \emph{Nature Physics}, vol.~8, no.~4, pp. 277--284, 2012.

\bibitem{bloch2008many}
I.~Bloch, J.~Dalibard, and W.~Zwerger, ``Many-body physics with ultracold
  gases,'' \emph{Reviews of modern physics}, vol.~80, no.~3, p. 885, 2008.

\bibitem{cerezo2020variational}
M.~Cerezo, A.~Poremba, L.~Cincio, and P.~J. Coles, ``Variational quantum
  fidelity estimation,'' \emph{Quantum}, vol.~4, p. 248, 2020.

\bibitem{cirac1995quantum}
J.~I. Cirac and P.~Zoller, ``Quantum computations with cold trapped ions,''
  \emph{Physical review letters}, vol.~74, no.~20, p. 4091, 1995.

\bibitem{cramer2010efficient}
M.~Cramer, M.~B. Plenio, S.~T. Flammia, R.~Somma, D.~Gross, S.~D. Bartlett,
  O.~Landon-Cardinal, D.~Poulin, and Y.-K. Liu, ``Efficient quantum state
  tomography,'' \emph{Nature communications}, vol.~1, no.~1, p. 149, 2010.

\bibitem{de2015fast}
A.~De, ``Fast quantum control for weakly nonlinear qubits: On two-quadrature
  adiabatic gates,'' \emph{arXiv preprint arXiv:1509.07905}, 2015.

\bibitem{de2010selective}
P.~C. De~Groot, J.~Lisenfeld, R.~N. Schouten, S.~Ashhab, A.~Lupa{\c{s}}cu,
  C.~J. P.~M. Harmans, and J.~E. Mooij, ``Selective darkening of degenerate
  transitions demonstrated with two superconducting quantum bits,''
  \emph{Nature Physics}, vol.~6, no.~10, pp. 763--766, 2010.

\bibitem{earnest2021pulse}
N.~Earnest, C.~Tornow, and D.~J. Egger, ``Pulse-efficient circuit transpilation
  for quantum applications on cross-resonance-based hardware,'' \emph{Physical
  Review Research}, vol.~3, no.~4, p. 043088, 2021.

\bibitem{egger2023study}
D.~J. Egger, C.~Capecci, B.~Pokharel, P.~K. Barkoutsos, L.~E. Fischer,
  L.~Guidoni, and I.~Tavernelli, ``A study of the pulse-based variational
  quantum eigensolver on cross-resonance based hardware,'' \emph{arXiv preprint
  arXiv:2303.02410}, 2023.

\bibitem{englund2005controlling}
D.~Englund, D.~Fattal, E.~Waks, G.~Solomon, B.~Zhang, T.~Nakaoka, Y.~Arakawa,
  Y.~Yamamoto, and J.~Vu{\v{c}}kovi{\'c}, ``Controlling the spontaneous
  emission rate of single quantum dots in a two-dimensional photonic crystal,''
  \emph{Physical review letters}, vol.~95, no.~1, p. 013904, 2005.

\bibitem{gambetta2012characterization}
J.~M. Gambetta, A.~D. C\'orcoles, S.~T. Merkel, B.~R. Johnson, J.~A. Smolin,
  J.~M. Chow, C.~A. Ryan, C.~Rigetti, S.~Poletto, T.~A. Ohki, M.~B. Ketchen,
  and M.~Steffen, ``Characterization of addressability by simultaneous
  randomized benchmarking,'' \emph{Phys. Rev. Lett.}, vol. 109, p. 240504,
  2012.

\bibitem{gambetta2011analytic}
J.~M. Gambetta, F.~Motzoi, S.~T. Merkel, and F.~K. Wilhelm, ``Analytic control
  methods for high-fidelity unitary operations in a weakly nonlinear
  oscillator,'' \emph{Physical Review A}, vol.~83, no.~1, p. 012308, 2011.

\bibitem{giurgica2020digital}
T.~Giurgica-Tiron, Y.~Hindy, R.~LaRose, A.~Mari, and W.~J. Zeng, ``Digital zero
  noise extrapolation for quantum error mitigation,'' in \emph{2020 IEEE
  International Conference on Quantum Computing and Engineering (QCE)}.\hskip
  1em plus 0.5em minus 0.4em\relax IEEE, 2020, pp. 306--316.

\bibitem{gross2017quantum}
C.~Gross and I.~Bloch, ``Quantum simulations with ultracold atoms in optical
  lattices,'' \emph{Science}, vol. 357, no. 6355, pp. 995--1001, 2017.

\bibitem{hanson2007spins}
R.~Hanson, L.~P. Kouwenhoven, J.~R. Petta, S.~Tarucha, and L.~M.~K.
  Vandersypen, ``Spins in few-electron quantum dots,'' \emph{Reviews of modern
  physics}, vol.~79, no.~4, p. 1217, 2007.

\bibitem{huang2022learning}
H.-Y. Huang, ``Learning quantum states from their classical shadows,''
  \emph{Nature Reviews Physics}, vol.~4, no.~2, pp. 81--81, 2022.

\bibitem{huang2020predicting}
H.-Y. Huang, R.~Kueng, and J.~Preskill, ``Predicting many properties of a
  quantum system from very few measurements,'' \emph{Nature Physics}, vol.~16,
  no.~10, pp. 1050--1057, 2020.

\bibitem{huang2022provably}
H.-Y. Huang, R.~Kueng, G.~Torlai, V.~V. Albert, and J.~Preskill, ``Provably
  efficient machine learning for quantum many-body problems,'' \emph{Science},
  vol. 377, no. 6613, p. eabk3333, 2022.

\bibitem{imamog1999quantum}
A.~Imamog, D.~D. Awschalom, G.~Burkard, D.~P. DiVincenzo, D.~Loss, M.~Sherwin,
  A.~Small \emph{et~al.}, ``Quantum information processing using quantum dot
  spins and cavity qed,'' \emph{Physical review letters}, vol.~83, no.~20, p.
  4204, 1999.

\bibitem{jaksch2005cold}
D.~Jaksch and P.~Zoller, ``The cold atom hubbard toolbox,'' \emph{Annals of
  physics}, vol. 315, no.~1, pp. 52--79, 2005.

\bibitem{johansson2012qutip}
J.~R. Johansson, P.~D. Nation, and F.~Nori, ``Qutip: An open-source python
  framework for the dynamics of open quantum systems,'' \emph{Computer Physics
  Communications}, vol. 183, no.~8, pp. 1760--1772, 2012.

\bibitem{kandala2019error}
A.~Kandala, K.~Temme, A.~D. C{\'o}rcoles, A.~Mezzacapo, J.~M. Chow, and J.~M.
  Gambetta, ``Error mitigation extends the computational reach of a noisy
  quantum processor,'' \emph{Nature}, vol. 567, no. 7749, pp. 491--495, 2019.

\bibitem{kjaergaard2020superconducting}
M.~Kjaergaard, M.~E. Schwartz, J.~Braum{\"u}ller, P.~Krantz, J.~I.-J. Wang,
  S.~Gustavsson, and W.~D. Oliver, ``Superconducting qubits: Current state of
  play,'' \emph{Annual Review of Condensed Matter Physics}, vol.~11, pp.
  369--395, 2020.

\bibitem{krantz2019quantum}
P.~Krantz, M.~Kjaergaard, F.~Yan, T.~P. Orlando, S.~Gustavsson, and W.~D.
  Oliver, ``A quantum engineer's guide to superconducting qubits,''
  \emph{Applied physics reviews}, vol.~6, no.~2, p. 021318, 2019.

\bibitem{larose2022mitiq}
R.~LaRose, A.~Mari, S.~Kaiser, P.~J. Karalekas, A.~A. Alves, P.~Czarnik,
  M.~El~Mandouh, M.~H. Gordon, Y.~Hindy, A.~Robertson \emph{et~al.}, ``Mitiq: A
  software package for error mitigation on noisy quantum computers,''
  \emph{Quantum}, vol.~6, p. 774, 2022.

\bibitem{leibfried2003quantum}
D.~Leibfried, R.~Blatt, C.~Monroe, and D.~Wineland, ``Quantum dynamics of
  single trapped ions,'' \emph{Reviews of Modern Physics}, vol.~75, no.~1, p.
  281, 2003.

\bibitem{lewenstein2007ultracold}
M.~Lewenstein, A.~Sanpera, V.~Ahufinger, B.~Damski, A.~Sen, and U.~Sen,
  ``Ultracold atomic gases in optical lattices: mimicking condensed matter
  physics and beyond,'' \emph{Advances in Physics}, vol.~56, no.~2, pp.
  243--379, 2007.

\bibitem{Li2022pulselevelnoisy}
\BIBentryALTinterwordspacing
B.~Li, S.~Ahmed, S.~Saraogi, N.~Lambert, F.~Nori, A.~Pitchford, and N.~Shammah,
  ``Pulse-level noisy quantum circuits with {Q}u{T}i{P},'' \emph{{Quantum}},
  vol.~6, p. 630, Jan. 2022. [Online]. Available:
  \url{https://doi.org/10.22331/q-2022-01-24-630}
\BIBentrySTDinterwordspacing

\bibitem{li2017efficient}
Y.~Li and S.~C. Benjamin, ``Efficient variational quantum simulator
  incorporating active error minimization,'' \emph{Physical Review X}, vol.~7,
  no.~2, p. 021050, 2017.

\bibitem{liang2022pan}
Z.~Liang, J.~Cheng, H.~Ren, H.~Wang, F.~Hua, Y.~Ding, F.~Chong, S.~Han, Y.~Shi,
  and X.~Qian, ``Pan: Pulse ansatz on nisq machines,'' \emph{arXiv preprint
  arXiv:2208.01215}, 2022.

\bibitem{liang2023towards}
Z.~Liang, J.~Cheng, Z.~Song, H.~Ren, R.~Yang, H.~Wang, K.~Liu, P.~Kogge, T.~Li,
  Y.~Ding \emph{et~al.}, ``Towards advantages of parameterized quantum
  pulses,'' \emph{arXiv preprint arXiv:2304.09253}, 2023.

\bibitem{liang2021can}
Z.~Liang, Z.~Wang, J.~Yang, L.~Yang, Y.~Shi, and W.~Jiang, ``Can noise on
  qubits be learned in quantum neural network? a case study on quantumflow,''
  in \emph{2021 IEEE/ACM International Conference On Computer Aided Design
  (ICCAD)}.\hskip 1em plus 0.5em minus 0.4em\relax IEEE, 2021, pp. 1--7.

\bibitem{magesan2011scalable}
E.~Magesan, J.~M. Gambetta, and J.~Emerson, ``Scalable and robust randomized
  benchmarking of quantum processes,'' \emph{Physical review letters}, vol.
  106, no.~18, p. 180504, 2011.

\bibitem{mckay2018qiskit}
D.~C. McKay, T.~Alexander, L.~Bello, M.~J. Biercuk, L.~Bishop, J.~Chen, J.~M.
  Chow, A.~D. C{\'o}rcoles, D.~Egger, S.~Filipp \emph{et~al.}, ``Qiskit backend
  specifications for openqasm and openpulse experiments,'' \emph{arXiv preprint
  arXiv:1809.03452}, 2018.

\bibitem{mckay2019three}
D.~C. McKay, S.~Sheldon, J.~A. Smolin, J.~M. Chow, and J.~M. Gambetta,
  ``Three-qubit randomized benchmarking,'' \emph{Physical review letters}, vol.
  122, no.~20, p. 200502, 2019.

\bibitem{meitei2021gate}
O.~R. Meitei, B.~T. Gard, G.~S. Barron, D.~P. Pappas, S.~E. Economou,
  E.~Barnes, and N.~J. Mayhall, ``Gate-free state preparation for fast
  variational quantum eigensolver simulations,'' \emph{npj Quantum
  Information}, vol.~7, no.~1, p. 155, 2021.

\bibitem{motzoi2009simple}
F.~Motzoi, J.~M. Gambetta, P.~Rebentrost, and F.~K. Wilhelm, ``Simple pulses
  for elimination of leakage in weakly nonlinear qubits,'' \emph{Physical
  review letters}, vol. 103, no.~11, p. 110501, 2009.

\bibitem{petta2005coherent}
J.~R. Petta, A.~C. Johnson, J.~M. Taylor, E.~A. Laird, A.~Yacoby, M.~D. Lukin,
  C.~M. Marcus, M.~P. Hanson, and A.~C. Gossard, ``Coherent manipulation of
  coupled electron spins in semiconductor quantum dots,'' \emph{Science}, vol.
  309, no. 5744, pp. 2180--2184, 2005.

\bibitem{preskill2018quantum}
J.~Preskill, ``Quantum computing in the nisq era and beyond,'' \emph{Quantum},
  vol.~2, p.~79, 2018.

\bibitem{proctor2022measuring}
T.~Proctor, K.~Rudinger, K.~Young, E.~Nielsen, and R.~Blume-Kohout, ``Measuring
  the capabilities of quantum computers,'' \emph{Nature Physics}, vol.~18,
  no.~1, pp. 75--79, 2022.

\bibitem{rigetti2010fully}
C.~Rigetti and M.~Devoret, ``Fully microwave-tunable universal gates in
  superconducting qubits with linear couplings and fixed transition
  frequencies,'' \emph{Physical Review B}, vol.~81, no.~13, p. 134507, 2010.

\bibitem{ryan2017hardware}
C.~A. Ryan, B.~R. Johnson, D.~Rist{\`e}, B.~Donovan, and T.~A. Ohki, ``Hardware
  for dynamic quantum computing,'' \emph{Review of Scientific Instruments},
  vol.~88, no.~10, p. 104703, 2017.

\bibitem{sheldon2016procedure}
S.~Sheldon, E.~Magesan, J.~M. Chow, and J.~M. Gambetta, ``Procedure for
  systematically tuning up cross-talk in the cross-resonance gate,''
  \emph{Physical Review A}, vol.~93, no.~6, p. 060302, 2016.

\bibitem{smith2022summary}
K.~N. Smith, G.~S. Ravi, T.~Alexander, N.~T. Bronn, A.~Carvalho,
  A.~Cervera-Lierta, F.~T. Chong, J.~M. Chow, M.~Cubeddu, A.~Hashim
  \emph{et~al.}, ``Summary: Chicago quantum exchange (cqe) pulse-level quantum
  control workshop,'' \emph{arXiv preprint arXiv:2202.13600}, 2022.

\bibitem{temme2017error}
K.~Temme, S.~Bravyi, and J.~M. Gambetta, ``Error mitigation for short-depth
  quantum circuits,'' \emph{Physical review letters}, vol. 119, no.~18, p.
  180509, 2017.

\bibitem{wang2022quest}
H.~Wang, P.~Liu, J.~Cheng, Z.~Liang, J.~Gu, Z.~Li, Y.~Ding, W.~Jiang, Y.~Shi,
  X.~Qian \emph{et~al.}, ``Quest: Graph transformer for quantum circuit
  reliability estimation,'' \emph{arXiv preprint arXiv:2210.16724}, 2022.

\bibitem{yu2022statistical}
X.-D. Yu, J.~Shang, and O.~G{\"u}hne, ``Statistical methods for quantum state
  verification and fidelity estimation,'' \emph{Advanced Quantum Technologies},
  vol.~5, no.~5, p. 2100126, 2022.

\end{thebibliography}
%%%%%%%%%%%%%%%%%%%%%%%%%%%%%%%%%%%%

\end{document}